\documentclass[draft]{agujournal2019}
\usepackage{url} 
\usepackage{lineno}
\usepackage[inline]{trackchanges} 
\usepackage{soul}
\usepackage{amsmath}
\usepackage{siunitx}
\draftfalse

\journalname{JGR: Earth Surface}

\begin{document}

%
%


\title{A micro-continuum physics-based model for cohesive sediment gravity flows across mudslide, mudflow, and turbidity current regimes}

%
%




\authors{Mitchell D. Jans\affil{1}, Cyprien Soulaine\affil{3}, Judy Q. Yang\affil{4,5}, and Ian C. Bourg\affil{1,2}}

\affiliation{1}{Department of Civil and Environmental Engineering, Princeton University, Princeton University, Princeton, NJ, USA}
 \affiliation{2}{High Meadows Environmental Institute, Princeton University, Princeton University, Princeton, NJ, USA}
 \affiliation{3}{Institut des Sciences de la Terre d'Orléans, UMR 7327, Univ Orleans, CNRS, BRGM, OSUC, Orleans, France}
 \affiliation{4}{Department of Civil, Environmental, and Geo-Engineering, University of Minnesota Twin Cities, Minneapolis, Minnesota, USA}
 \affiliation{5}{St. Anthony Falls Laboratory, University of Minnesota Twin Cities, Minneapolis, Minnesota, USA}





\correspondingauthor{=Mitchell D. Jans=}{=mitchelljans@princeton.edu=}



\begin{keypoints}
\item A model based on the Darcy-Brinkman-Biot Framework was developed for cohesive sediment gravity flows and validated with experiments.
\item The computational fluid dynamics model reproduced flow regimes and maximum flow speeds.
\item Both intrinsic and extrinsic sediment properties can exert first-order controls on flow speed and flow morphology.
\end{keypoints}
\textbf{Keywords}: Darcy-Brinkman-Biot, Sediment Transport, Sediment Gravity Flows, Clay Minerals, Debris Flows, Turbidity Currents \\
\textbf{Index Terms}: 3022, 1815, 1861, 1952, 4558
%
%

%
%


\begin{abstract}

Gravity driven sediment flows are responsible for a major portion of sediment redistribution within oceans, reservoirs, and lakes, with
important implications in coastal erosion, siltation, carbon burial, and contaminant migration in aquatic systems.
Despite the ubiquity of this phenomenon, current mechanistic understanding of sediment gravity flows (SGFs) remains limited. This
knowledge gap is particularly acute in the case of cohesive, fine-grained sediments (i.e., muds) due to the complex properties of the clay
matrix, including low permeability, viscoplastic rheology, and flocculation. In this work, we develop a computational fluid dynamics
model that accurately predicts key features of cohesive, clay-rich SGFs based on independent measurements of the relation between sediment solid fraction and
rheological yield stress. In particular, the model captures the four primary flow regimes (low density turbidity currents, high density turbidity currents, mudflows, and mudslides)
observed in lock-exchange experiments with slurries containing smectite or kaolinite clay. The model is validated through comparison
with previous experimental observations of sediment flow morphology, speed, and runout distance. Overall, we demonstrate the ability to predict the influence of intrinsic (particle size, grain density, and rheology) and extrinsic sediment properties (sediment topography and solid fraction) in the development of self-sustaining cohesive SGFs.
\end{abstract}

\section*{Plain Language Summary}
Predicting where and when large-scale sediment flows occur underwater has important implications in our understanding of geophysical hazards, pollutant transport, and how underwater landscapes evolve. Currently, there is limited predictive ability to understand underwater sediment flows that contain significant quantities of clay (i.e., mud). In this work, we develop computer simulations that can predict the macroscopic flow behavior of clay-rich sediment flows, including how fast they travel and major flow features. We identify key characteristics that differ between clay types (bentonite vs. kaolinite) and demonstrate how they impact flow properties. Our results suggest the importance of the time-dependence of these properties in achieving accurate prediction of mudflow runout distances. Overall, this work advances our understanding of how clay-rich sediment moves underwater.

%
%

%


%
%
%
%

\section{Introduction}
Gravity driven sediment flows, often referred to as sediment gravity flows (SGFs), are dense, sediment-laden currents that move down slope under their own weight \cite{Dasgupta2003}. They constitute one of the largest sediment transport processes globally, carrying more sediment than all rivers combined, and are among the most consequential subaqueous geophysical processes \cite{Clarke1990,Hampton1996,Jerolmack2019,Voigtlander2024}. These flows are notable for their ability to achieve self-sustaining currents for hundreds of kilometers traveling at speeds that can exceed 20 $m\,s^{-1}$ on the ocean floor \cite{Bottner2024}. They are ubiquitous across subaqueous environments and while marine environments host the most extensive SGFs, similar processes occur in lakes and reservoirs, where they efficiently redistribute sediment across basins, influencing stratigraphy and long-term sediment budgets \cite{Meiburg2010,Chamoun2016}.

A recurrent theme in studies of cohesive SGFs is the need to understand the mechanisms and governing physics behind these flows. Improved mechanistic understanding is essential to our ability to predict, and eventually mitigate, the impacts of these flows on marine infrastructure (oil and gas operations, communication cables, dredging operations) \cite{Locat2002}, contaminant fate and transport \cite{Pohl2020}, deep ocean carbon fluxes \cite{Galy2007}, siltation of reservoirs and lakes \cite{Schleiss2016}, and paleoceanographic reconstructions \cite{Hesse2006}. Progress in developing a predictive understanding of these flows has been hindered by their sensitivity to a variety of intrinsic and extrinsic sediment properties including particle size, density, mineralogy, rheology, sediment topography, and biological activity \cite{Jing2018,Craig2019}.

The present study investigates SGFs formed by cohesive clay-laden sediments. Although granular, non-cohesive SGFs have been extensively studied, the transition between granular SGFs and cohesive SGFs occurs at sufficiently low clay solid fraction [reported below 5\% \cite{Hampton1972,Baas2002}] that clay-related cohesion within the flow can be significant even in relatively coarse-grained sediments \cite{Dasgupta2003}. With the principal abiotic cohesive force in most surficial sediments being due to colloidal interactions between clay minerals, detailed understanding of how these minerals impact SGFs is essential \cite{Marr2001,Zhou2025}. For example, clay particle interactions have been demonstrated to increase the stability of the resulting debrite, attenuate turbulence after incipient motion begins, and increase the likelihood for hydroplaning of the flow head due to low permeability which inhibits sediment consolidation \cite{Du2022}. Although cohesive sediments exhibit a range of compositions, for simplicity, we focus on sediment rich in clay minerals, i.e., layered silicate minerals with small size and large specific surface area \cite{Sposito1999}. This focus is justified as mud, which consists of clay minerals and organic matter, is the most abundant sediment on the ocean floor \cite{Velde1995} and globally, clay minerals constitute roughly 50\% of all sediment mass \cite{Ito2017}. Cohesive clay SGFs are expected to exhibit sensitivity to clay mineralogy, clay fraction, and aqueous chemistry due to the impact of these features on clay colloidal interactions at the microscopic scale which can affect macroscale SGF regime development \cite{Zadehali2025,Zheng2025}.

To contextualize the behavior of SGFs, a range of SGF regime classification schemes have been proposed, each dependent on differing flow characteristics \cite{Baas2009,Hermidas2018}. Here we utilize a scheme in which SGF regimes are broadly categorized by flow morphology and the primary mechanism that supports sediment within the flow. The origin of this classification framework can be traced to \citeA{Middleton1973} in which four primary types of SGFs were defined: turbidity currents, fluidized sediment flow, grain flow, and debris flow. For clay-rich SGFs, turbidity currents and debris flows are the primary flow regimes, as fluidized sediment flow is restricted by the clay's low permeability and grain flow does not occur due to the small sediment size and negligible nature of direct frictional forces between the grains \cite{Guven1992}. 

As in \citeA{Baker2017}, we partition the turbidity current and debris flow categories into two regimes each: low- and high-density turbidity currents, and mudflows and mudslides. This leaves clay-rich SGFs with four distinct regimes: low-density turbidity currents (LDTC), high-density turbidity currents (HDTC), mudflows (MF), and mudslides (MS). Turbidity currents are the most common SGF and are characterized by the presence of turbulence throughout the flow, low sediment fractions, quasi-Newtonian rheology and abundant Kelvin-Helmholtz instabilities above the body of the flow. They occur at dilute sediment fractions, where flow matrix strength is low to nonexistent and flow behavior is driven by turbulence. Consistent with \citeA{Baker2017}, we define HDTC by the presence of significant gradients in sediment solid fraction within the flow, as opposed to the more uniform solid fractions seen in LDTC.

The remaining two regimes are classified as debris flows due to the importance of the matrix strength resulting from cohesive forces within the flow \cite{Iverson1997}. These regimes are distinguished from turbidity currents by their coherent laminar flow behavior. Mudslides are the most viscous form of SGFs and occur at elevated sediment fractions with significant cohesion between sediment particles. They are distinguished from mudflows by their low flow speed, minimal internal deformation, small runout distances, and absence of hydroplaning of the flow head. One important consideration that can cause classification difficulties in field studies is that transition between regimes can occur mid-flow due to changes in sediment solid fraction or slope \cite{Mohrig2003,Felix2006,Talling2013}; however, such transitions did not occur in the simulated flows presented here. 

Because of the episodic, underwater nature of SGFs, direct observations in natural systems have been historically limited, though methodological advances are beginning to alleviate this observational gap \cite{Paull2018,Hage2019,Talling2023}. As a consequence, experimental understanding of SGFs is based predominantly on post-event examination of field sites \cite{Stevenson2018,Bottner2024}. These are supplemented by so-called "lock-exchange" experiments where well-characterized sediment-water mixtures are released into a flume and driven downstream by the density difference between the mixture and water \cite{Haza2013,Craig2019,Sobocinska2022}. Although these studies have improved our understanding, the experiments are limited in size due to laboratory constraints, which raises unanswered questions surrounding lab-field scalability. Field studies avoid the scalability concerns, but they generally lack information on pre-event conditions or data collection during flow events. These constrains warrant the development of methods to enable direct multiscale observation of SGFs allowing for easier data collection during flow events \cite{Choi2024}. 

One promising approach for advancing fundamental understanding of SGFs is the development of physics-based numerical models \cite{Talling2015}. In particular, computational fluid dynamics (CFD) simulations have been widely used to investigate geophysical flows including avalanches, sand dune migration, and lahars \cite{Giri2006,Oda2011,Mead2017}. A notable strength of CFD simulations is their ability to examine flow conditions and coupled physical processes that are difficult or impossible to isolate experimentally and to evaluate how these interactions give rise to emergent large-scale behavior. Previous CFD studies of SGFs have considered a wide range of sediment types, both non-cohesive and cohesive, using various modeling approaches, including Eulerian and Lagrangian representations of sediment \cite{Rzadkiewicz1997,Chauchat2017,Guo2024,Chauchat2017,Peng2025}. These studies have explored the role of turbulence closures and rheological formulations using reduced-complexity representations such as volume-of-fluid (VOF)-based approaches \cite{VonBoetticher2016}. However, while VOF methods are effective for capturing bulk flow geometry and interface evolution, they are less suited to resolving solid fraction evolution and fluid-solid momentum exchange within a porous or depositional bed. In contrast, a micro-continuum Euler-Euler framework, such as that adopted here, explicitly represents fluid flow within the sediment, making it more appropriate for investigating the internal mechanics of cohesive SGFs \cite{Lee2019,Guo2023}.

In this study, we develop a micro-continuum open-source CFD model that incorporates a shear-rate and solid fraction-dependent rheology (Herschel-Bulkley–Quemada) and enables predictive simulations across different types of clay minerals and sediment solid fractions. Our approach is based on the so-called Darcy-Brinkman-Biot (DBB) framework, previously developed to represent the ductile mechanics of clayey media, such as consolidation and desiccation cracking \cite{Carrillo2019,Carrillo2021}. The framework is tested here, for the first time, for its ability to predict large-scale clay deformation such as that associated with SGFs. Specifically, we evaluate its ability to predict the properties of cohesive SGFs formed by two major clay minerals: kaolinite and smectite (a.k.a. bentonite). We benchmark our simulations by predicting, with no fitting parameters, the results of flume lock-exchange experiments carried out by \citeA{Baker2017}. We then use the validated model to examine the influence of rheology, sediment fraction, sediment mineralogy, and sediment height on flow speed, runout distance, and flow morphology. This CFD study is the first, to our knowledge, to utilize independently measured sediment properties to predict the transition between SGF regimes, compare mineralogical influences, and capture maximum flow speeds from the microscale material properties of cohesive sediments. 
\section{Methods}
\subsection{Mathematical Framework}
Computational fluid dynamic simulations were conducted using an updated version of the Darcy-Brinkman-Biot solver, termed SedDBB, in OpenFOAM 7, a C++ library that uses the Finite Volume Method to discretize differential equations for computational fluid dynamics. This framework consists of four conservation equations for fluid and solid mass (\ref{Fluid_Mass} \& \ref{Solid_Mass}), and momentum (\ref{Fluid_Momentum} \& \ref{Solid_Momentum}).
\begin{equation}
\frac{\partial \phi_f}{\partial t}+ \nabla \cdot \left(\phi_f\boldsymbol{U}_f\right)=0
\label{Fluid_Mass}
\end{equation}

\begin{equation}
\frac{\partial\phi_f \rho_f \boldsymbol{U}_f}{\partial t}
+ \nabla\cdot\left(\rho_f\phi_f\boldsymbol{U}_f \boldsymbol{U}_f\right)
= \nabla\cdot S-\phi_f\nabla p + \phi_f \rho_f \boldsymbol{g}
- K_d (\boldsymbol{U}_f-\boldsymbol{U}_s)
\label{Fluid_Momentum}
\end{equation}

\begin{equation}
\frac{\partial \phi_s}{\partial t}+\nabla\cdot \left(\phi_s\boldsymbol{U}_s\right)=0
\label{Solid_Mass}
\end{equation}
\begin{equation}
\frac{\partial\phi_s \rho_s \boldsymbol{U}_s}{\partial t}+\nabla\cdot (\rho_s\phi_s\boldsymbol{U}_s \boldsymbol{U}_s)=\nabla\cdot \sigma+\phi_s\nabla\cdot {\tau^s} + \phi_s \rho_s \boldsymbol{g} + K_d (\boldsymbol{U}_f-\boldsymbol{U}_s)
\label{Solid_Momentum}
\end{equation}
 
 In Eqs. \ref{Fluid_Mass}-\ref{Solid_Momentum}, $\phi_f$ and $\phi_s$ are the fluid and solid volume fractions, $\boldsymbol{U}_f$ and $\boldsymbol{U}_s$ are the intrinsic velocity of the fluid and solid, $\rho_f$ and $\rho_s$ are the fluid and solid density, $S$ and $\sigma$ are the volume averaged viscous stress tensor of the fluid and solid, \textit{p} is the fluid pressure, $\boldsymbol{g}$ is gravity, $\tau^s$ is Terzaghi's effective stress tensor, and $K_d$ is the fluid-solid drag coefficient. The definition of $\tau^s$ used in Eq. \ref{Solid_Momentum} can include applied external stresses and solid swelling pressure, but these are both neglected in the present work, such that $\tau^s$ =$-pI$.

Details on the derivation of Eqs. \ref{Fluid_Mass}-\ref{Solid_Momentum} are provided in \citeA{Carrillo2019} and \citeA{Carrillo2021}. In short, Eqs. \ref{Fluid_Mass}-\ref{Fluid_Momentum} are based on the well known Darcy-Brinkman representation of fluid flow in systems that contain both free space and porous regions \cite{Neale1974,Soulaine2024}, while Eqs. \ref{Solid_Mass}-\ref{Solid_Momentum} implement the capacity for ductile deformation of the porous solid in a manner consistent with Eqs. \ref{Fluid_Mass}-\ref{Fluid_Momentum} and with Biot poromechanics. The DBB framework has been applied to a variety of systems with bulk fluid and deformable microporous regions including deformable soils and sediments, coastal barriers, fractured shale, and biofilms \cite{Carrillo2019,Carrillo2021b,Carrillo2021,Kurz2022,Zheng2025B}. 

Crucially, Eqs. \ref{Fluid_Mass}-\ref{Solid_Momentum} have shown the capacity to predict key properties of clay-rich soils, sediments, and sedimentary rocks, such as consolidation and desiccation cracking, by incorporating constitutive relations for the material properties of the clay matrix at the scale of the computational grid, including its complex rheology and low permeability. However, previous applications of this framework have established its accuracy only in conditions where the solid undergoes relatively minor deformation. Its ability to predict much larger deformations, such as those associated with SGFs, is not presently known and is examined in the present study. 
\subsection{Clay Material Properties}
\subsubsection{Rheology and Density}
Both clay types examined here are modeled as incompressible with densities as reported in \citeA{Baker2017} (displayed in table \ref{tab:table-01}). We note that the reported bentonite density (2,300 \si{\kilogram\per\meter\cubed}) is below the theoretical density (2,800 \si{\kilogram\per\meter\cubed}) based on the crystallographic unit cell structure of smectite clay minerals, the main component of bentonite. This discrepancy likely reflects the strong hygroscopic tendency of bentonite, whereby dry bentonite equilibrated with humid air can hold significant quantities of absorbed water \cite{Berend1995,Li2024}. Bulk water density used in our simulations reflects the saline water used in experiments. 

The rheological behavior of subaqueous geophysical flow has been represented using a variety of constitutive models including, viscoplastic, $\mu$ (I), and coulomb-type rheologies \cite{Ancey2007,Trujillo-Vela2022}. In the present work, as in previous applications of Eqs. \ref{Fluid_Mass}- \ref{Solid_Momentum}, clay rheology is modeled using a viscoplastic framework based on the Herschel-Bulkley (Eq. \ref{HB1} \& \ref{HB2}) and Quemada relations (Eq. \ref{Q1} \& \ref{Q2}).
\begin{equation}
\sigma=\mu_s^{eff} \left(\nabla U_s + \left(\nabla U_s\right)^T-\left(\frac{2}{3}\right) \nabla\cdot\left(U_s I\right)\right)
\label{HB1}
\end{equation}
\begin{equation}
\mu_s^{eff}= min\left(\mu_s^0, \left(\frac{\tau}
{\eta}\ +\mu_s \eta^{n-1}\right)\right)
\label{HB2}
\end{equation}
\begin{equation}
\tau=\tau_0 \left(\frac{\phi_s / \phi_s^{max}}{\left(1-\frac{\phi_s}{ \phi_s^{max}}\right)}\right)^D
\label{Q1}
\end{equation}
\begin{equation}
\mu_s= \frac{\mu_0}{\left(1 - \frac{\phi_s}{\phi_s^{max}} \right)^2}
\label{Q2}
\end{equation}

 The Herschel-Bulkley model accounts for the existence of a yield stress $\tau$ and the possibility of shear thinning (\textit{n} $<$ 1) or thickening (\textit{n} $>$ 1) behavior at high shear rates $\eta$, phenomena well documented in clayey sediments \cite{Jeong2010,Jeong2012,Yang2014,Whorton2025}. In the present study, the yield stress $\tau$ was constrained by fitting Eq. {\ref{Q1}} (where  \textit{D} is a rheological scaling parameter and $\tau_0$ is a reference yield stress) to experiments carried out by \citeA{Baker2017}. The other parameters in Eq. \ref{HB2} were not determined experimentally. For simplicity, we set the flow index \textit{n} = 1, i.e. clay is modeled as a Bingham plastic. The values of $\mu_s^0$ and $\mu_s^0$ (the limiting viscosity at low and high shear rate) are also unconstrained experimentally in the conditions of the present work. They were set to values used in our previous work for clay-like media, informed by experiments on other clayey materials \cite{Coussot1995,Spearman2017}. Sensitivity of model predictions to these parameters is evaluated in Section \ref{SectionRheologySensitivity}. We note that while the use of a maximum viscosity $\mu_s^0$ is consistent with the known creep deformation of clayey media, in the conditions of this work, it mostly serves to ensure numerical stability, a practice commonly implemented to allow for a computationally feasible timestep \cite{VonBoetticher2016}. 

\begin{figure}[!htpb]
    \centering
    \includegraphics[width=\linewidth]{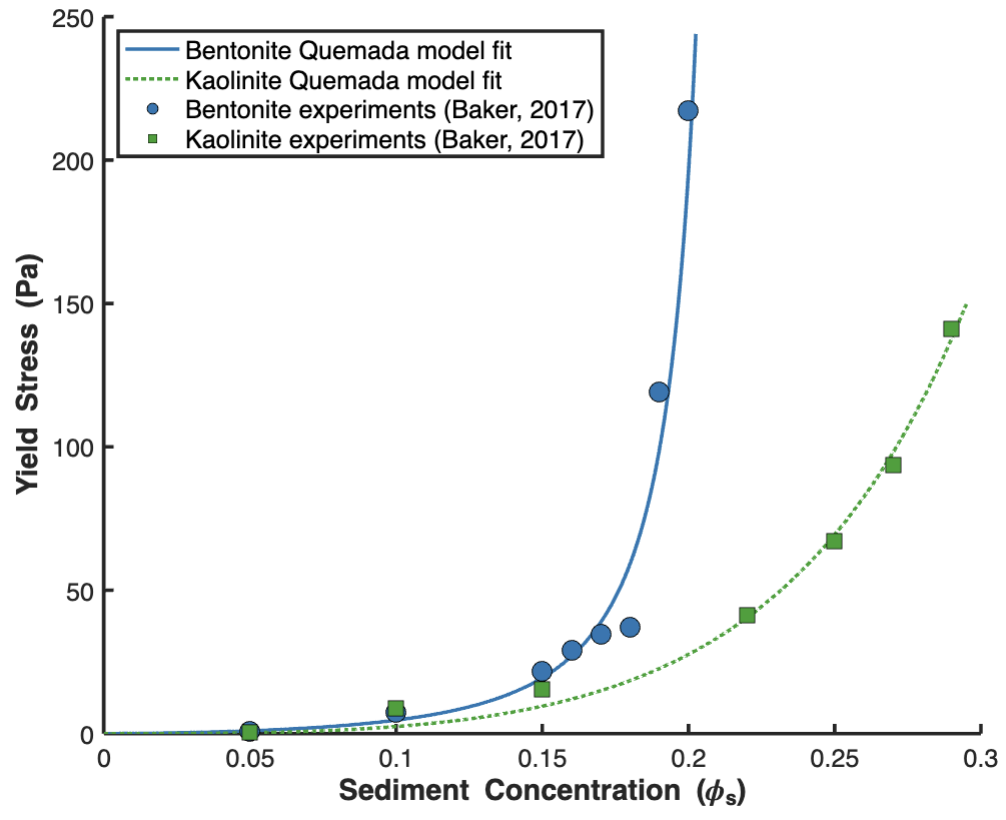}
    \caption{Yield stress $\tau$ as a function of sediment fraction ($\phi_s$). Symbols show experimental measurements by \citeA{Baker2017}. Lines represent best fits using the Quemada yield stress relation (Eq. \ref{Q1}) for bentonite and kaolinite. Associated best-fit values of $\tau_0$, \textit{D}, and $\phi_s^{max}$ are shown in Table \ref{tab:table-02}} 
    \label{fig:YS_Fit}
\end{figure}

The Quemada relation accounts for the dependence of yield stress and viscosity on sediment fraction ($\phi_s$). When used in conjunction, Eqs. \ref{HB1}-\ref{Q2} reproduce experimental rheological findings by increasing yield stress and viscosity for elevated sediment fractions while allowing for nearly Newtonian rheology of sediment flows at dilute sediment fractions \cite{Coussot1995,Pignon1997}. Values of $\tau_0$, \textit{D}, and $\phi_s^{max}$ were fit to rheological measurements reported by \citeA{Baker2017} using a bounded nonlinear least-square regression method. The resulting fits of yield stress $\tau$ as a function of sediment volume fraction $\phi_s$ are shown in Fig. \ref{fig:YS_Fit}. 
\begin{table}[!htpb]
    \caption{Rheological parameters used in the Herschel-Bulkley (Eq. \ref{HB1}-\ref{HB2}) and Quemada relations (Eqn. \ref{Q1}-\ref{Q2}). The first three parameters ($\tau_0$, \textit{D}, and $\phi_s^{max}$) were fitted to rheology measurements reported by \citeA{Baker2017} (Fig. \ref{fig:YS_Fit}). The other two parameters (\textit{n}, $\nu_0$) were set to generic values and have minimal impacts on SGF predictions as discussed in Section \ref{SectionRheologySensitivity}}
    \label{tab:table-02}
    \centering
    \begin{tabular}{l c c c}
        \hline
        \textbf{Parameter} & \textbf{Symbol} & \textbf{Value} & \textbf{Units} \\
        \hline
        Bentonite yield stress        & $\tau_{0,b}$      & 6.199   & Pa \\
        Kaolinite yield stress        & $\tau_{0,k}$      & 246.831 & Pa \\
        Bentonite fractal dimension   & $D_b$             & 1.5     & --  \\
        Kaolinite fractal dimension   & $D_k$             & 2.7     & --  \\
        Bentonite max volume fraction & $\phi_{s,b}^{\max}$ & 0.22  & --  \\
        Kaolinite max volume fraction & $\phi_{s,k}^{\max}$ & 0.65  & --  \\
        Bentonite flow index          & $n_b$             & 1       & --  \\
        Kaolinite flow index          & $n_k$             & 1       & --  \\
        Bentonite Maximum Viscosity          & $\nu_{b0}$             & 1       & $m^2 s^{-1}$  \\
        Kaolinite Maximum Viscosity         & $\nu_{k0}$            & 1       & $m^2 s^{-1}$ \\
        \hline
    \end{tabular}
\end{table}
\\

\subsubsection{Permeability and Drag }
The interphase momentum transfer term is modeled using the Ergun equation (Eq. \ref{ErgunEqu}), where $Re_p$ is the particle Reynolds number: 
\begin{equation}
K_d=\phi_s\mu_s\left(150\left(\frac{\phi_s}{\phi_f}\right)+1.75Re_p\right).
\label{ErgunEqu}
\end{equation}
At low $Re_p$, Eq. \ref{ErgunEqu} converges to the Kozeny-Carman relation, which was utilized in previous applications of Eqs \ref{HB1}-\ref{Q2} \cite{Enwald1996,Carrillo2019}. In the Ergun drag model, a particle size is utilized in conjunction with porosity ($\phi_f$) to calculate an effective permeability. As discussed in Section \ref{SectionDragSens}, the friction coefficient $K_d$ plays a dominant role in controlling fluid flow within the sediment bed and the settling velocity of suspended sediment. We note that the effective particle sizes used in this study (5.6 and 9.1 $\mu$m), taken from \citeA{Baker2017}, are significantly larger than the sizes of primary clay particles \cite{Mesri1971}, suggesting that these effective particle sizes represent the size of clay aggregates or flocs.

\subsection{Simulation Numerics and Algorithm}
Implementation of the SedDBB Model utilizes the twoPhaseEulerFoam solver architecture from OpenFOAM 7.0 to effectively resolve the four conservation equations noted above. While keeping the general multiphase Euler structure, substantial modifications were implemented in the code to improve computational efficiency and numerical accuracy while following the DBB framework. 

Consistent with previous DBB solvers, the solution is performed using the PISO (Pressure-Implicit-Split-Operator) algorithm \cite{Issa1986}. The PISO algorithm allows for accurate transient calculations. Timestep is adaptive and is constrained using the CFL (Courant-Friedrichs-Lewy) and Fourier numbers. These numbers ensure numerical stability for advection and momentum diffusive processes. The required timestep is generally imposed by the Fourier (momentum diffusion) number early in the simulation due to high sediment viscosity within the flow and low flow speeds. If momentum diffusion-based timestep limits are not properly imposed, a significant delay in flow initiation and a warping of the shape of the front can occur, leading to erroneous observations including lack of convergence with computational mesh refinement. The inclusion of a timestep constraint based on Fourier number resolves this issue as shown in the Supplementary Information (Fig. S1).

Departing from previous DBB solvers, the fluid and sediment momentum equations are coupled together in the velocity-pressure coupling operation \cite{Soulaine2015}. This allows for semi-implicit handling of drag, an important upgrade from previous solvers in systems with high drag. Additionally, conservation equations are
phase-averaged (intrinsic), in contrast with past DBB solvers \cite{Carrillo2021}.

\subsection{Simulation Setup}
\begin{figure}[!htpb]
    \centering
    \includegraphics[width=\linewidth]{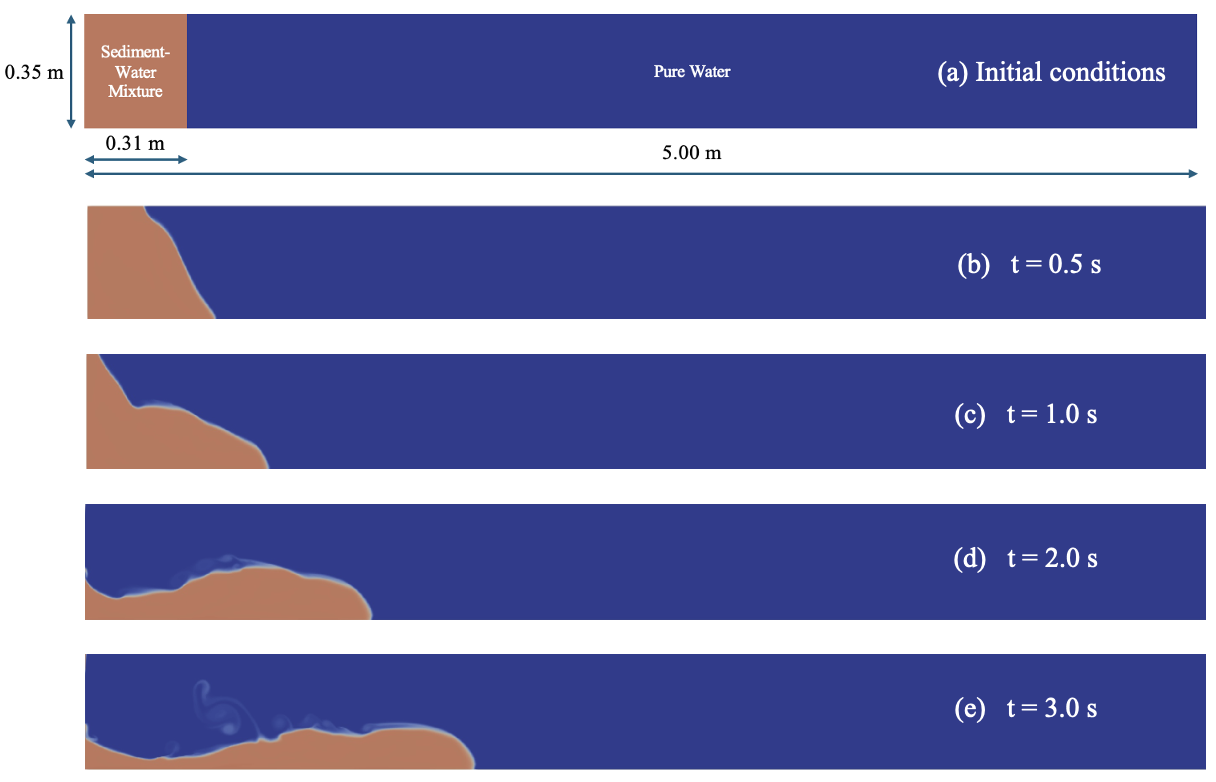}
    \caption{(\textbf{Top panel}) Schematic of the starting configuration of the lock-exchange experimental setup used by \citeA{Baker2017}. (\textbf{Lower panels}) Illustration of the simulated evolution of solid volume fraction ($\phi_s$) distribution in a representative simulated SGF. (Schematic dimensions not to scale).}
    \label{fig:Schematic}
\end{figure}
Our baseline simulations are intended to replicate the lock-exchange experiments documented in \citeA{Baker2017}. Sediments consisted of uniform mixtures of clay (either bentonite or kaolinite) and saltwater at solid volumetric fractions ranging from 1\% to 29\%. \citeA{Baker2017} analyzed their results to determine the influence of clay type and solid fraction on SGF flow speed, morphology, and eventual sediment runout distance. 

Our simulations represent the experimental flume shape (5.0 m length, 0.35 m depth, and 0.2 m width) using a two-dimensional grid (2000 x 140 cells) with dimensions of 5.0 m by 0.35 m. Further grid refinement had no impact to flow head velocity, as seen in Supplementary Information Fig. S2. The starting configuration consists of two distinct regions: the initial sediment-water mixture region (tan colored in Fig. \ref{fig:Schematic}), with dimensions of 0.35 m by 0.31 m, which holds an initial solid volume fraction $\phi_s$ $>$ 0, and the bulk water region, which is the remainder of the simulated domain that initially contains only water ($\phi_s$ = 0).

Boundary conditions are chosen to mimic the experimental setup by enforcing a slip boundary at the upper surface and no-slip conditions at the solid flume walls (i.e. at the lateral and lower boundaries). A subset of simulations were carried out using an explicit third dimension. Theses showed minimal influence on flow behavior, so all results reported here are from two-dimensional runs. 
  
\begin{table}[!htpb]
\caption{Model parameters used in the base-case simulations, based on \citeA{Baker2017}.}
\label{tab:table-01}
\centering
\begin{tabular}{l c c c}
\hline
\textbf{Parameter} & \textbf{Symbol} & \textbf{Value} & \textbf{Units}\\
\hline
Bentonite grain density  & $\rho_{sb}$ & 2,300                & kg\,m$^{-3}$ \\
Kaolinite grain density  & $\rho_{sk}$ & 2,600                & kg\,m$^{-3}$ \\
Water density               & $\rho_{w}$  & 1,027                & kg\,m$^{-3}$ \\
Water kinematic viscosity   & $\nu_{w}$   & $1.004\times10^{-6}$ & m$^{2}$\,s$^{-1}$ \\
Bentonite particle diameter & $d_{pb}$    & $5.6\times10^{-6}$   & m \\
Kaolinite particle diameter & $d_{pk}$    & $9.1\times10^{-6}$   & m \\
\hline
\end{tabular}
\end{table}

\subsection{Characterization of Simulation Results}
The main flow characteristics reported by \citeA{Baker2017} consist of the maximum speed of the SGF, its morphology, and its runout distance. The latter property was determined only for flows that stopped before reaching the end of the flume. All three properties require the use of a threshold $\phi_s$ value to delineate the shape of the sediment bed. In the present work, to determine flow speed and runout distances, the $\phi_s$ threshold value was set to 70\% of the initial sediment fraction in the sediment reservoir (i.e. for a 5\% initial solid fraction, a threshold of 3.5\% is utilized). Results showed minimal sensitivity to the choice of threshold value as the leading head of the flow largely remained well-defined. Horizontal locations of the sediment flow front were calculated every 0.1 s with flow speed calculated for the corresponding intervals. The maximum flow speed corresponds to the 0.1 s interval during which the sediment front showed the most horizontal displacement. Figure \ref{fig:Schematic} illustrates the temporal evolution of a representative SGF, including its initial acceleration phase and transition to sustained flow speed. 

In determining the flow regimes (LDTC, HDTC, MF and MS), classifications were identified based on flow head morphology, instabilities in the sediment-water interface, and trailing sediment deposits. Morphological characteristics used in SGF regime identification were based on descriptions in \citeA{Baker2017}. 
\section{Experimental Validation}
\begin{figure*}[t]
    \centering
    \includegraphics[width=\textwidth]{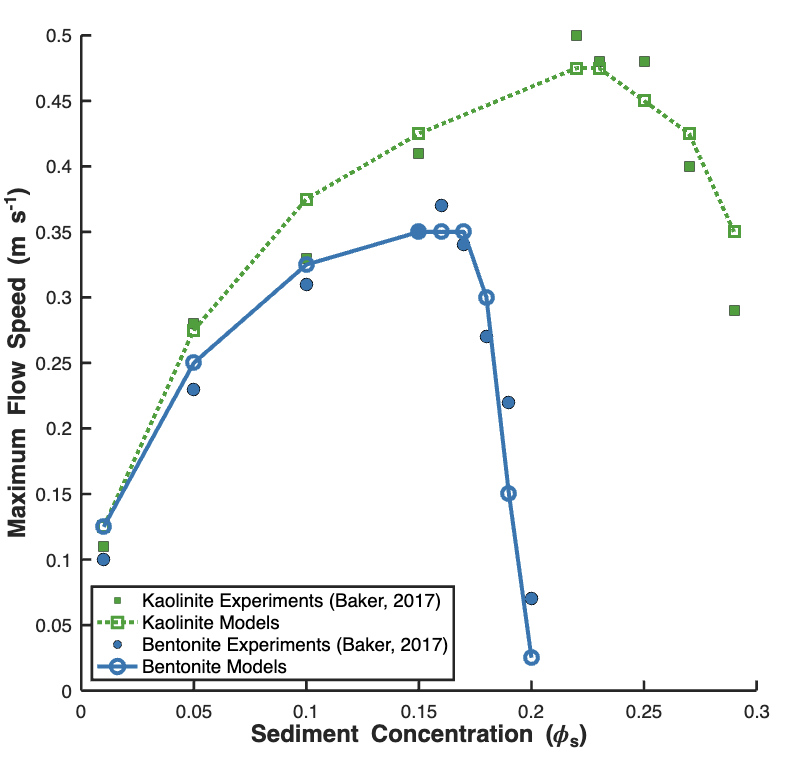}
    \caption{Comparison of predicted and measured maximum flow speed of the SGF head. Filled symbols indicate experimental results, solid lines with open symbols denote model predictions. The lines connecting simulation predictions are drawn to guide the eye. Blue color corresponds to bentonite, green to kaolinite.}
    \label{fig:ComparingSpeed}
\end{figure*}

Despite its lack of fitting parameters, the model accurately replicated experimental results on the maximum head flow speed as a function of sediment mineralogy and solid fraction reported by \citeA{Baker2017} (Fig. \ref{fig:ComparingSpeed}). Predicted maximum flow speeds are consistent with experimental values for both dilute and concentrated SGFs. In particular, the model accurately predicts the non-monotonic relationship between maximum flow speed and sediment solid fraction and the consistently faster flow of kaolinite relative to bentonite sediments. 

\begin{figure}[!htpb]
    \centering
    \hspace*{-0.15\linewidth}
    \includegraphics[width=1.3\textwidth]{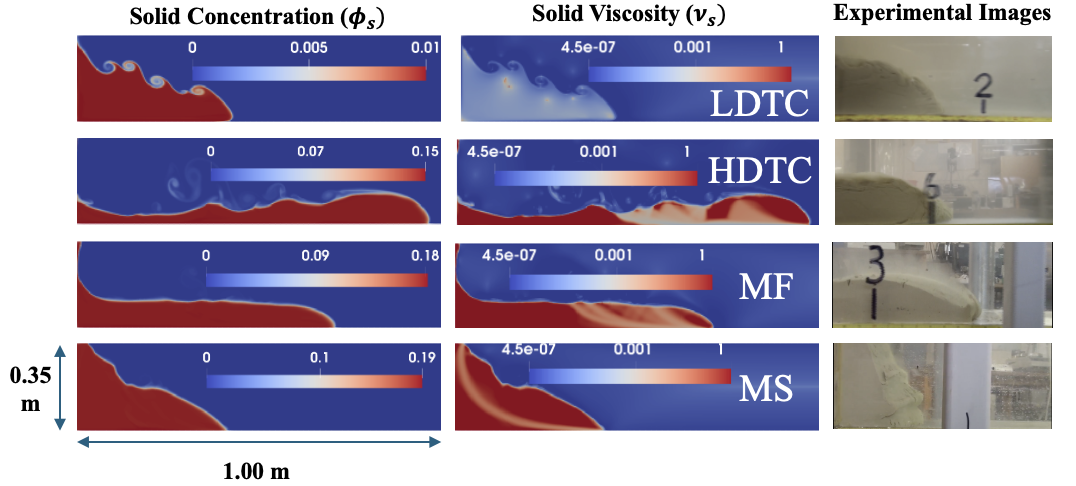}
    \caption{(\textbf{Left}) Simulation visualization showing the spatial distribution of solid fraction in representative examples of the LDTC, HDTC, MF, and MS regimes at \textit{t} = 3.5 seconds for bentonite simulations with initial solid fraction $\phi_s$ = 1\%, 15\%, 18\%, and 19\% (from top to bottom). (\textbf{Center}) Predicted solid kinematic viscosity fields in the same conditions represented using a logarithmic scale in the range of 4.5$\times10^{-7}$ to 1 $m^2s^{-1}$. (\textbf{Right}) Images courtesy of Megan Baker (Durham U.) based on the experiments described in \citeA{Baker2017} showing the four SGF regimes.}
    \label{fig:tcanther}
\end{figure}

\begin{figure*}[t]
    \centering
    \includegraphics[width=\textwidth]{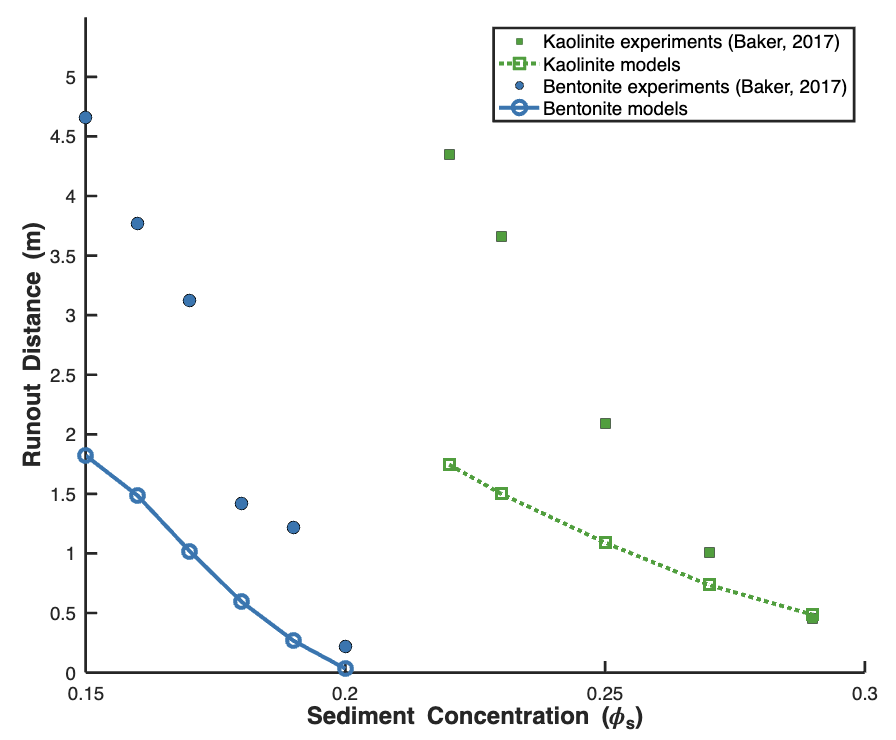}
    \caption{Comparison of predicted and measured maximum runout distance. Note only experimental runs that had a quantifiable runout distance are included. Filled symbols indicate experimental finding, solid line with open symbols denote modeled output. The lines connecting simulation predictions are drawn to guide the eye. Blue corresponds to bentonite, green to kaolinite}
    \label{fig:ComparingRunout}
\end{figure*}

The model also accurately predicts the expected SGF regime transitions across the range of sediment fractions shown in Fig. \ref{fig:tcanther}. Characteristic LDTC behavior, with a well-mixed sediment distribution within the flow and pronounced Kelvin-Helmholtz instabilities, is observed in the most dilute cases. With increasing solid fraction, the simulations also predict HDTC behavior, with a dense lower layer and a dilute turbulent upper layer; MF behavior, with suppressed turbulence but sustained bulk motion and elevated velocities; and MS behavior, with limited internal deformation and a wedge shaped front. The only significant inconsistency between simulation predictions and experimental results is that runout distances are systematically under predicted (Fig. \ref{fig:ComparingRunout}). Hypotheses that may explain this disparity are discussed in Section \ref{ImproveRunout}.

\begin{figure}[!htpb]
    \centering
    \hspace*{-0.15\linewidth}
    \includegraphics[width=1.3\textwidth]{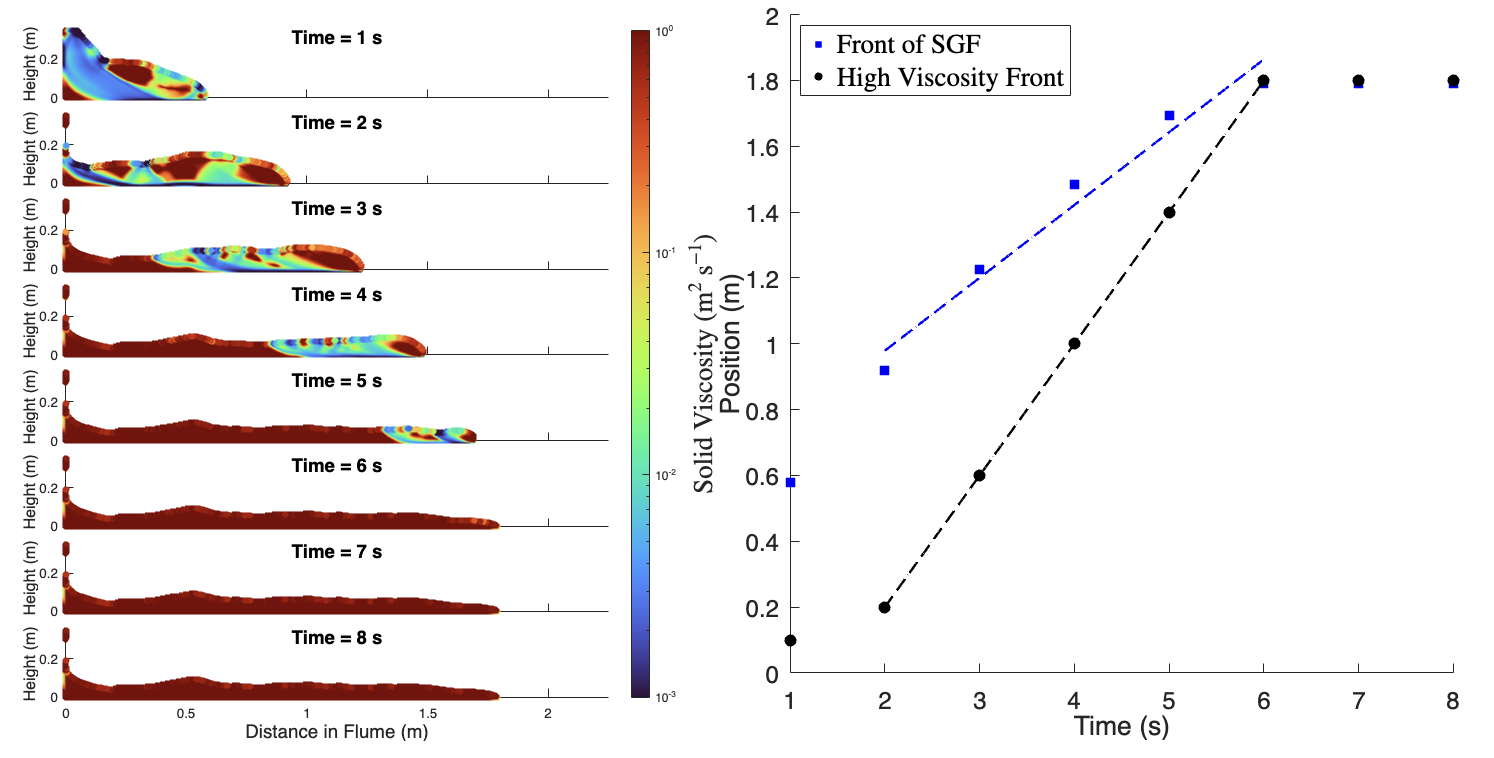}
    \caption{Time evolution of solid viscosity in a representative bentonite flow with $\phi_s$= 16\%. (\textbf{Left}) Logarithmic visualization of solid kinematic viscosity values between $10^{-3}$ and 1 $m^2s^{-1}$ throughout the flow. (\textbf{Right}) SGF flow front and highly viscous ($\nu$=1 $m^2 s^{-1}$) front through time. Linear fit based on positions between 2 and 6 seconds shown in dashed lines. The convergence of the two lines between \textit{t} = 6 and 7 s coincides with the cessation of flow occurs.}.
    \label{fig:Thix}
\end{figure}

As a corollary to the flow speeds and runout distances reported in Figs. \ref{fig:ComparingSpeed} and \ref{fig:ComparingRunout}, we evaluated the mechanism associated with SGF flow and flow cessation by visualizing the spatial and temporal evolution of predicted sediment viscosity in our baseline simulation. The results are presented as logarithmic-scale viscosity maps in Fig. \ref{fig:Thix}. Results suggest that flow is highly sensitive to the shear-rate dependence of viscosity (Eq. \ref{HB2}) as shown by the large viscosity gradients within the flow, including the formation of a high viscosity plug near the flow front sliding on a low viscosity lubrication layer at the base of the flow. Figure \ref{fig:Thix} further demonstrates that the cessation of flow manifests itself as a sharp viscosity front that initiates at the trailing edge of the flow and that propagates forward at a constant speed. Abrupt cessation of flow occurs when this high viscosity front overtakes the SGF flow front.
\section{Sensitivity Analysis}
This section uses the solver developed in this work to systematically evaluate the sensitivity of SGFs to key sediment characteristics. This is carried out through a series of tests in which most model parameters are fixed at the baseline values listed in Table \ref{tab:ModelEval} unless otherwise noted. Rheological parameters are set to the bentonite values listed in Table \ref{tab:table-01}. 
\begin{table}
\caption{Model parameters used as the base case for all sensitivity analyses}
\label{tab:ModelEval}
\centering
\begin{tabular}{l c c c}
\hline
\textbf{Parameter} & \textbf{Symbol} & \textbf{Value} & \textbf{Units} \\
\hline
Sediment Grain Density      & $\rho_{s}$ & 2,300              & kg\,m$^{-3}$ \\
Initial Solid Fraction& $\phi_{s}$ & 0.10               &              \\
Effective Particle Diameter& k         & $5.6\times10^{-6}$ & m            \\
Sediment Height       & h          & 0.35               & m            \\
Sediment Viscosity& $\mu$      & Bentonite          &              \\
\hline
\end{tabular}
\end{table}
\subsection{Sensitivity to Solid Grain Density}
\begin{figure}[!htpb]
    \centering
    \hspace*{-0.15\linewidth}
    \includegraphics[width=1.3\textwidth]{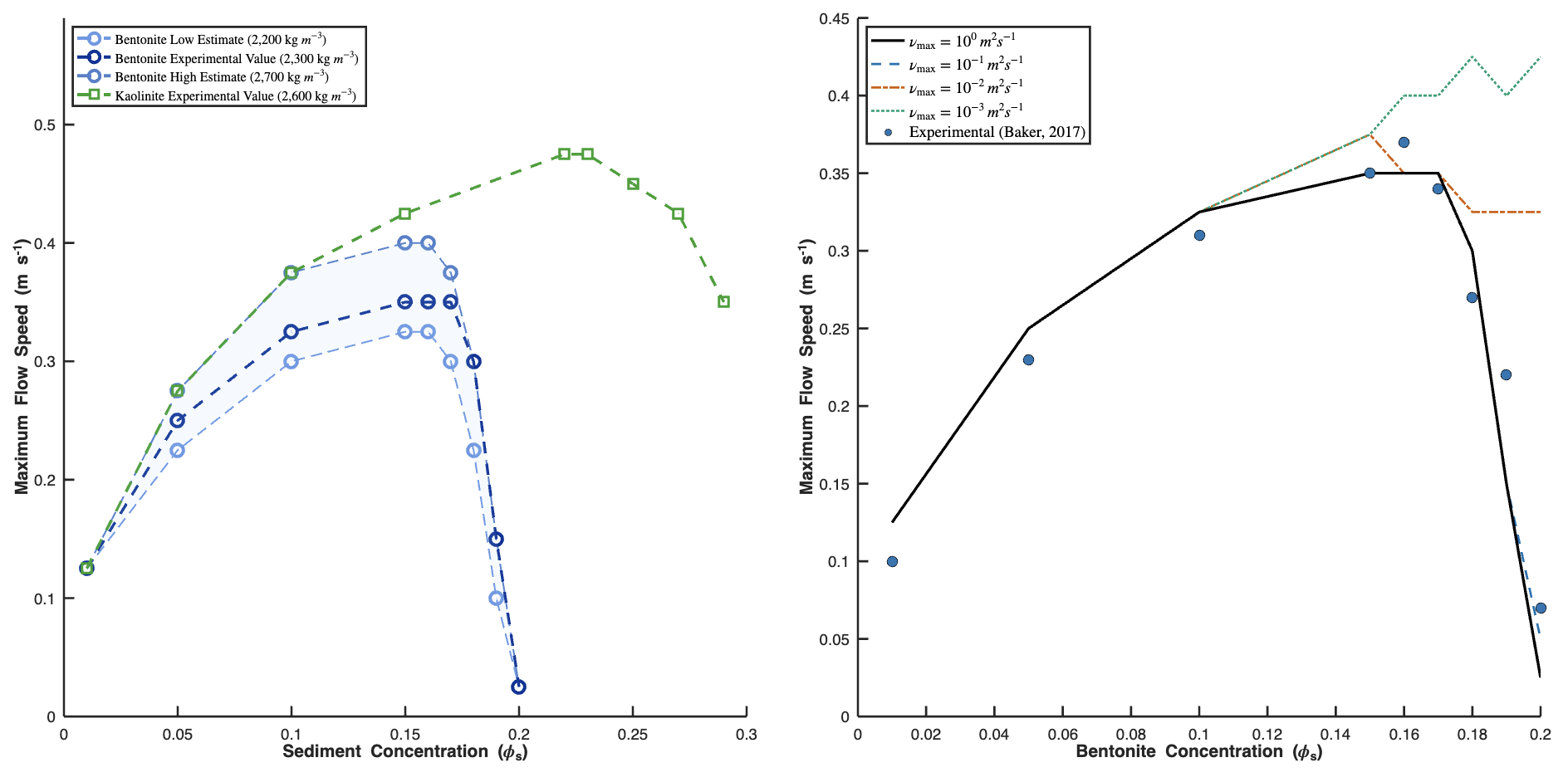}
    \caption{ (\textbf{Left}) Sensitivity of predicted maximum flow speed to sediment dry grain density $\rho_s$ for bentonite SGFs at varying sediment solid fractions. Lines represent simulations based on the range of bentonite grain densities reported in previous studies of cohesive sediments (roughly 2,200 to 2,700 kg m$^{-3}$). The influence of $\rho_s$ on flow speed varies with sediment fraction, indicating that the importance of dry grain density estimation increases with $\phi_s$ in conditions with limited plastic matrix strength (LDTC, HDTC) before decreasing at higher solid fractions (MF, MS). (\textbf{Right}) Sensitivity of predicted maximum flow speed to maximum solid viscosity ($\mu_s^0$ in Eq. \ref{HB2}, expressed here as $\nu_{max}$) for bentonite SGFs at varying sediment fraction compared with experiments. Model run markers are omitted due to significant overlap between points.}
    \label{fig:SedViscosityDensity}
\end{figure}

The density contrast between water and the sediment-laden flow is a primary control on SGF motion. Literature values indicate that bentonite has a measured grain density ranging from 2,200 to 2,700 $kg\, m^{-3}$ owing to differences in mineralogy and density estimation approaches \cite{Mehta2022}. Uncertainty in sediment grain density is therefore an important source of uncertainty in predicted maximum flow speeds. To isolate the effect of density differences associated with clay mineralogy, sediment density was varied systematically as a model input parameter. As shown in Fig. \ref{fig:SedViscosityDensity}, maximum flow speed increases monotonically with grain density. Across the tested range, variations in assumed grain density produce differences in flow speed of up to approximately 30\%, demonstrating that uncertainty in sediment density can substantially affect predicted SGF mobility. This effect may also help explain the slightly higher flow speeds of kaolinite relative to bentonite, even when their yield stress values are comparable at lower solid fractions. 
\subsection{Sensitivity to Rheological Parameters}
\label{SectionRheologySensitivity}
Although several rheological parameters were constrained by the yield stress measurements reported by \citeA{Baker2017}, three parameters used in Eqs. \ref{HB1}-\ref{Q2} were not constrained: the flow index \textit{n} and the limiting viscosity values at very high and low shear rates, $\mu_0$ and $\mu^0_s$. To evaluate model sensitivity to these rheological parameters, we modified their values over a wide range. 

Flow index is known to vary between clay sediments. In particularly, shear thinning behavior (\textit{n} $<$ 1) is well documented in rheological characterizations of clayey sediments \cite{Whorton2025}. Neverless, in field-scale studies the corresponding Herschel-Bulkley rheology is routinely simplified to a Bingham plastic rheology (\textit{n} = 1). Results presented in Fig. S3, may explain this apparent discrepancy, as the value of \textit{n} (within a range of 0.4 to 1.6) has essentially no impact on the speed of the front in the conditions of this work. This is likely due to the limited shear rate achieved in these lock-exchange experiments. In larger scale flows, in which faster speeds are reached, higher shear rates are expected, particularly at the base of the flow, such that the value of \textit{n} may impact SGF mobility in geologic settings. 

We also evaluated the impact of minimum ($\mu_0$) and maximum ($\mu_s^0$) sediment viscosities on simulation results. Sensitivity to $\mu_0$ was minimal in the expected use range of water-like viscosities and maximum flow speed was only impacted at values above $10^{-4}$ m$^2$ s$^{-1}$  as shown in Supplementary Information S4. In the case of $\mu_s^0$, results obtained by varying its value over three orders of magnitude (from $10^{-3}$ to 1 m$^2$ s$^{-1}$) are shown in Fig. \ref{fig:SedViscosityDensity}. Results at $\phi_s$ $<$ 0.15 (in the turbidity current regime) are insensitive to $\mu_s^0$, likely due to the low yield stress of dilute sediments. This indicates that $\mu^0_s$ in this regime can be chosen based on numerical considerations (where lower values of $\mu^0_s$ accelerate convergence). 

In contrast, MF and particularly MS regimes are sensitive to maximum viscosity values. As seen in Fig. $\ref{fig:SedViscosityDensity}$, as $\phi_s$ approaches 20 \%, predicted maximum flow speeds begin to significantly diverge. This behavior is consistent with the expectation that $\mu^0_s$ controls the creep deformation of the material (i.e. its deformation at stresses below the yield stress) which becomes the predominant mode of sediment deformation in the MS regime. 
\subsection{Sensitivity to the Choice of Drag Model}
\label{SectionDragSens}
\begin{figure*}[!htpb]
    \centering
    \hspace*{-0.15\linewidth}
    \includegraphics[width=1.3\textwidth]{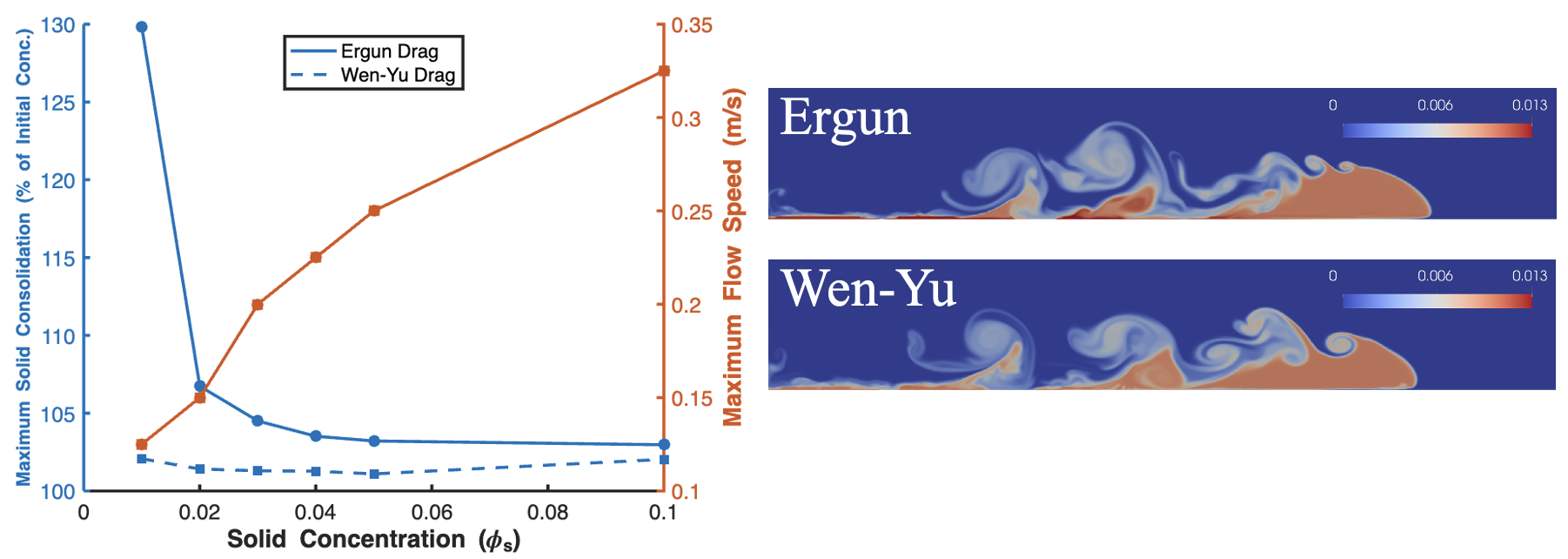}
    \caption{(\textbf{Left}) Influence of drag formulation on sediment deposition and flow speed. Blue symbols (with lines drawn to guide the eye) show the predicted maximum solid fraction within the flow after 10 s for simulations using the Ergun and Wen–Yu drag models across a range of initial solid fractions. Values greater than 100\% reflect the gravitational deposition of sediment at the lower boundary. Red symbols, plotted relative to the right axis, show the corresponding maximum flow speeds (lines obtained with different drag model overlap). While the choice of drag formulation has minimal influence on maximum flow speed, it strongly affects sediment deposition at low initial fractions. (\textbf{Right}) Flow visualizations at $\phi_s$= 0.01 illustrate this behavior, with the Ergun drag model producing greater sediment deposition (darker red color near the lower boundary) than the Wen–Yu formulation.}
    \label{fig:ParticleSizeRunout}
\end{figure*}

One of the defining characteristics of cohesive SGFs is their fine particle size resulting from the predominance of clay minerals. This small size ensures high fluid-solid friction within the flow and very low particle settling velocities. This limiting of settling within the flow has been demonstrated to prolong peak velocity values and maintain high sediment fractions in turbidity currents \cite{He2018}. In this section, we examine flow velocity and sediment consolidation dependence on drag through altering sediment particle size and interphase drag models. From our baseline simulation (i.e, bentonite at $\phi_s= 10\%$), we vary sediment effective particle size between 2 $\mu$m and 80 $\mu$m. In addition, we carry out additional simulations to compare the results obtained with the Ergun and Wen-Yu drag relations for dilute bentonite fractions ranging from 1\% to 10\% \cite{Enwald1996}.
As shown in Fig. \ref{fig:ParticleSizeRunout}, the choice of interphase drag model has little influence on maximum flow speed, but it does affect sediment gravitational settling at dilute sediment fractions and the development of interfacial instabilities at the water–flow boundary. At 1\% bentonite, the Ergun relation predicts significantly greater settling than the Wen--Yu relation, likely because the packed-bed assumptions underlying Ergun are less appropriate for such dilute suspensions and, therefore, may under-predict fluid-solid friction. This difference between drag models decreases substantially when $\phi_s \geq 5\%$ and becomes negligible at $\phi_s \geq$ 10\%. 

As shown in Fig. S5, at $\phi_s$ = 10\% and in the range of effective particle sizes reported by \citeA{Baker2017} for bentonite (5.6 $\mu$m) and kaolinite (9.1 $\mu$m), particle size has little influence on particle settling and essentially no impact on maximum flow speed. In contrast, particle sizes above approximately 20 $\mu$m lead to substantially greater deposition, consistent with reduced interphase drag and weaker fluid-sediment coupling. Despite this increase in deposition, the maximum flow speed remains nearly unchanged across particle sizes, likely because peak velocities occur shortly after flow release, before differences in settling become pronounced.

These results suggest that the maximum flow speed and flow morphology at clay fractions examined in this study are generally not highly sensitive to the choice of drag model, although drag formulation may be important for dilute flows, particularly when particle settling and runout distances are of interest. 

\subsection{Sensitivity to Initial Sediment Column Height}
The height of the initial sediment column also influences the velocity of SGFs due to its control of initial potential energy within the system. Owing to space limitations, lock-exchange experiments are often limited in their ability to reproduce flows driven by large differences in the height of the sediment surface and this has been noted as a significant limitation in scaling between field and lab studies \cite{Choi2024}. Here, we evaluate the importance of sediment column height through altering the initial sediment reservoir height between 0.05 m and 1 m. 

\begin{figure*}[!htpb]
    \centering
    \hspace*{-0.15\linewidth}
    \includegraphics[width=1.3\textwidth]{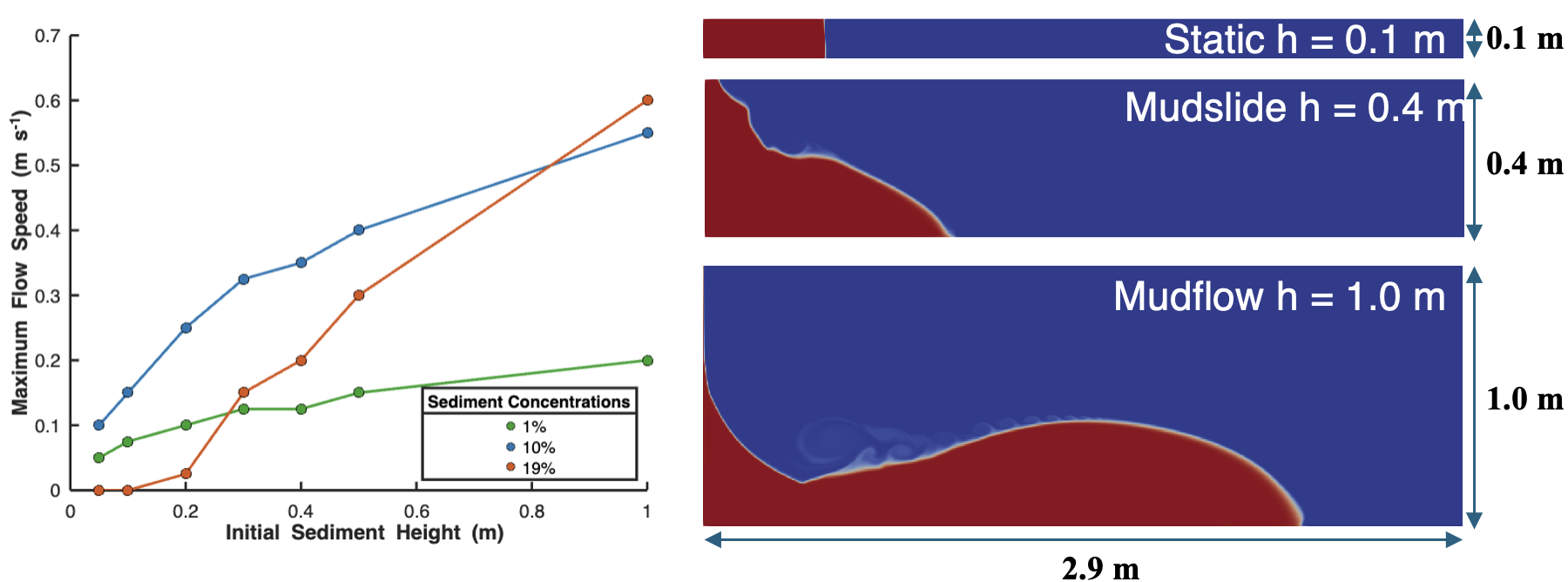}
    \caption{Influence of initial sediment height on maximum flow speed. (\textbf{Left}) Simulated maximum flow speeds for varying initial sediment heights for bentonite with initial $\phi_s$ = 0.01, 0.10, and 0.19. (\textbf{Right}) Visualization of $\phi_s$ = 0.19 flows at \textit{t} = 4 seconds for \textit{h} = 0.10 m, 0.40 m, 1.00 m. Flow morphologies demonstrate a transition (driven by initial sediment column height) between no failure, MS, and MF regimes.}
    \label{fig:SlopeImpact}
\end{figure*}

As seen in Fig.\ref{fig:SlopeImpact}, sediment column height shows a strong positive correlation with maximum speed for all SGF regimes. Additionally, height can be a critical factor in determining whether slope failure occurs as shown by the results at $\phi_s$= 19\%, where 0.05 and 0.10 m models remain immobile while taller initial sediment columns exceed failure thresholds and undergo mobilization. Changes in height can also directly influence SGF regimes with an increase in height altering flow regimes from MS to MF and eventually transitioning into turbidity currents at sufficiently high velocity flows. 

\section{Discussion}
\subsection{Improvement of the runout distance prediction}
\label{ImproveRunout}
\begin{figure*}[!htpb]
    \centering
    \hspace*{-0.15\linewidth}
    \includegraphics[width=1.3\textwidth]{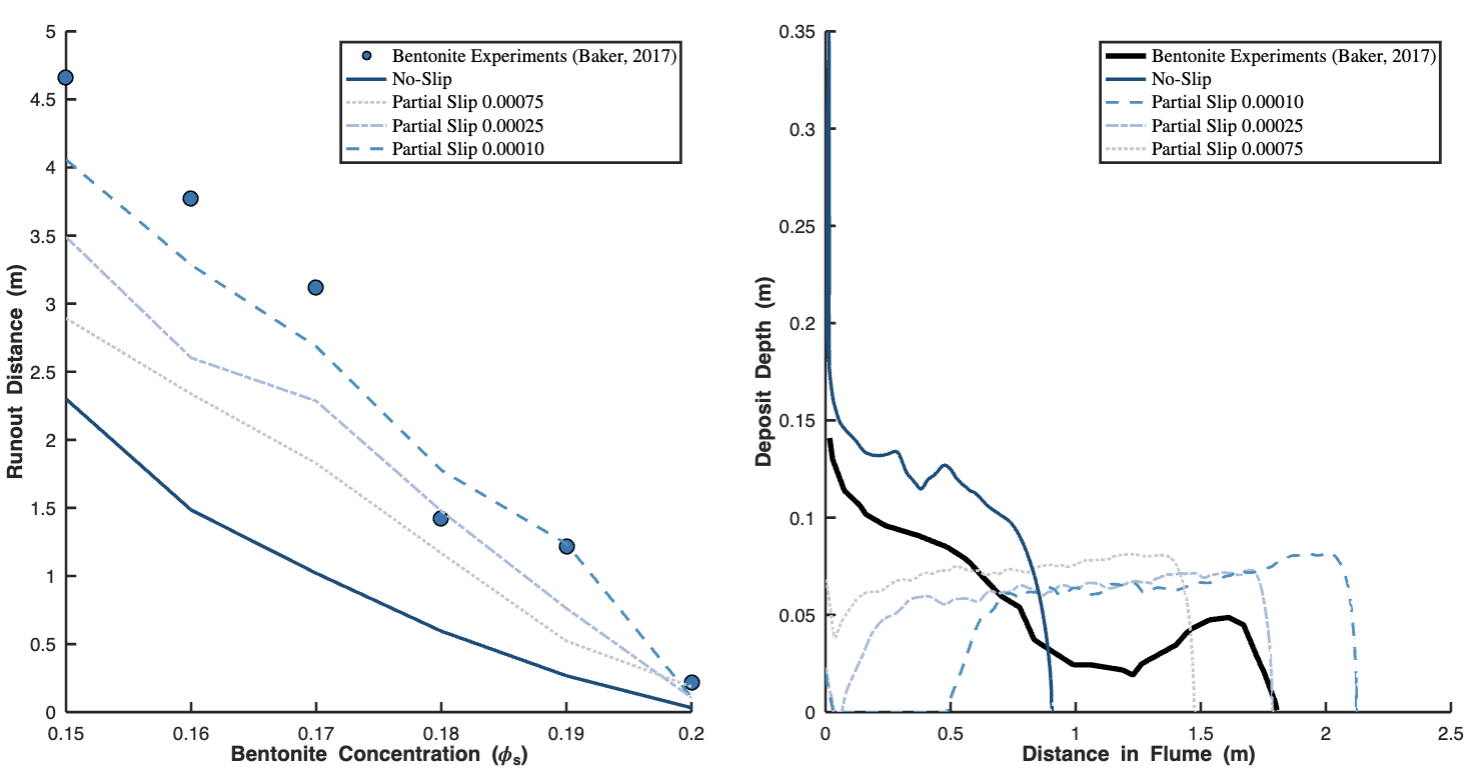}
    \caption{Sensitivity of simulation results to the use of a partial slip boundary condition at the flume walls. (\textbf{Left}) Comparison of simulated bentonite SGF runout distance with laboratory measurements \cite{Baker2017} as a function of wall boundary conditions. Partial slip coefficients reported in the legend reflect the fractional tendency of the sediment to stick at the solid walls (1 = no slip; 0 = full slip) (\textbf{Right}) Final morphology of the sediment deposit for 18\% bentonite flows. Simulations employing a no-slip boundary condition systematically underpredict runout distance across all solid fractions. Introducing a partial-slip boundary condition significantly increases runout distance, with a slip coefficient of about 0.0001 yielding closest agreement with experimental observations. However, the qualitative shape of eventual sediment deposits observed experimentally (dark blue in right panel) is significantly better predicted by the model with no slip (solid blue line) than by model predictions with partial slip (dashed lines).}
    \label{fig:PartialSlipImpact}
\end{figure*}

As shown in Fig. \ref{fig:ComparingRunout}, the model generally underpredicts runout distance relative to the experiments. We hypothesize three possible explanations for this discrepancy. First, the results suggest that the basal boundary condition plays an important role in controlling mobility. Specifically, the Herschel-Bulkley rheology of the sediment causes significant shear localization at the lower boundary of the simulated system (Fig. \ref{fig:Thix}), and the resulting lubrifying effect may be underestimated due to the finite grid resolution. Introducing a partial-slip condition improves agreement with the experiments in terms of runout distance, as seen in Fig. \ref{fig:PartialSlipImpact}, while leaving the maximum flow speed largely unchanged. However, this modification produces deposit morphologies that are inconsistent with the experiments, including lack of retention in the initial sediment region and an overall deposit that remains too thick. This indicates that although basal slip can improve bulk mobility metrics, it does not fully capture the depositional behavior of the flow. 

Second, we considered whether the imposed maximum viscosity $\mu^0_s$ may be limiting runout. While reducing $\mu^0_s$ can increase runout distance in some cases, such changes are generally accompanied by an a significantly worse prediction of SGF flow speed, as shown in Fig. \ref{fig:SedViscosityDensity}. This suggests that $\mu^0_s$ alone is unlikely to explain the runout discrepancy.

Third, we hypothesize that the model may be lacking an important time-dependent rheological control, particularly on the backside of the flow. Although the overall qualitative deposit shapes are broadly reasonable, the flows tend to arrest too early. One possible explanation is that the Herschel–Bulkley rheology model responds instantaneously to changes in shear rate, such that regions experiencing a reduction in shear rate, especially in the trailing portion of the flow, can rapidly regain high effective viscosity and prematurely ``lock up" as their shear rate decreases. In natural cohesive suspensions, viscosity has been observed to respond non-instantaneously to decreases in shear rate, possibly over timescale of seconds to minutes, due to the finite time-scale associated with clay microstructural evolution \cite{Pignon1998,vanKessel1998,Mewis2009}. This so-called thixotropic effect could play an important role in increasing SGF runout distance without impacting maximum flow speed as it directly impacts the abrupt deceleration phase of the flow but allows for accurate capturing of the initial acceleration by a Herschel-Bulkley rheological model \cite{Jeon2018}. In the context of the results shown in Fig. \ref{fig:Thix}, we hypothesize that thixotropy would decrease the speed of the high-viscosity front (black symbols in Fig. \ref{fig:Thix}b) and retard the time when this front overtakes the SGF flow front. We do not pursue a thixotropic rheological formulation here because it would substantially increase model complexity and would require additional experimental constraints that are not presently available in conditions matching the SGF observations of \citeA{Baker2017}. 
\subsection{Dimensionless Numbers and Impact of Yield Stress}
\begin{figure*}[!htpb]
    \hspace*{-0.15\linewidth}
    \includegraphics[width=1.3\textwidth]{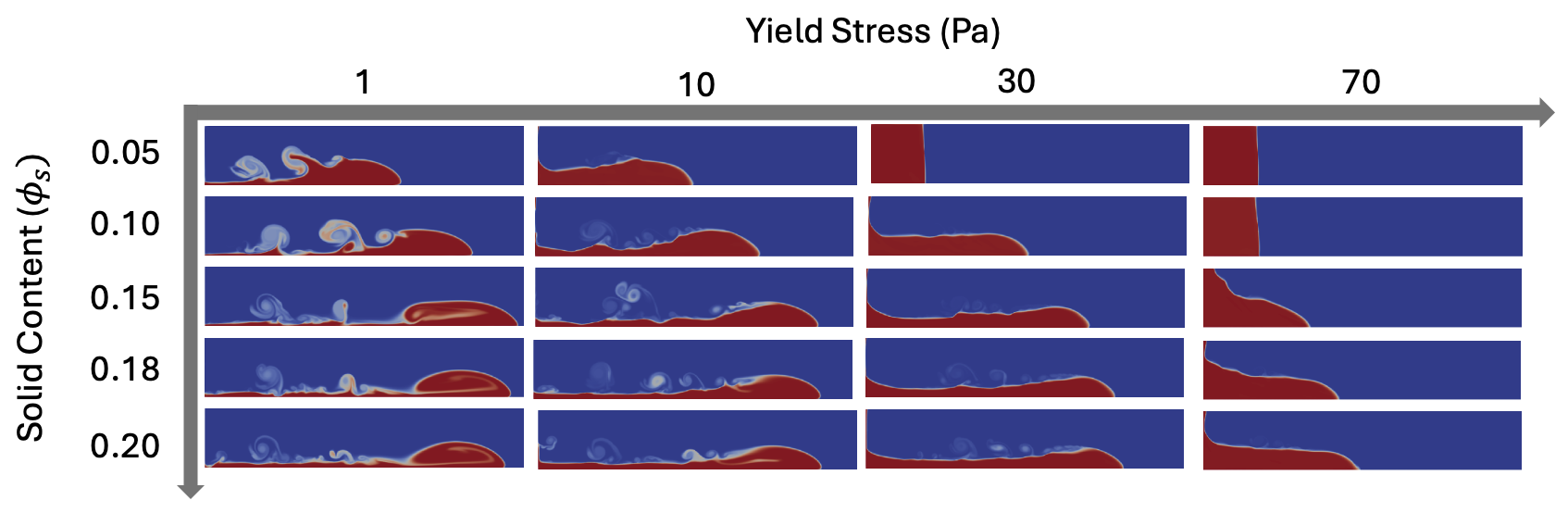}
  \caption{Snapshots of simulated cohesive SGFs at t = 4 s for varying solid volume fractions ($\phi_s$ = 0.05–0.20, rows) and yield stresses ($\tau$ = 1–70 Pa, columns). Color contours show the sediment phase distribution (red = high fraction, blue = zero fraction). Increasing solid fraction enhances flow coherence and front propagation, while increasing yield stress suppresses mobility, leading to shorter runout distances and thicker, more arrested deposits. Intermediate cases exhibit pronounced head–tail structure and Kelvin–Helmholtz–type instabilities at the surface of the sediment bed, particularly at lower yield stresses and moderate fractions.}
  \label{fig:YieldStressInfluence}
\end{figure*}
\begin{figure}[!htpb]
    \centering
    \hspace*{-0.15\linewidth}
    \includegraphics[width=1.3\textwidth]{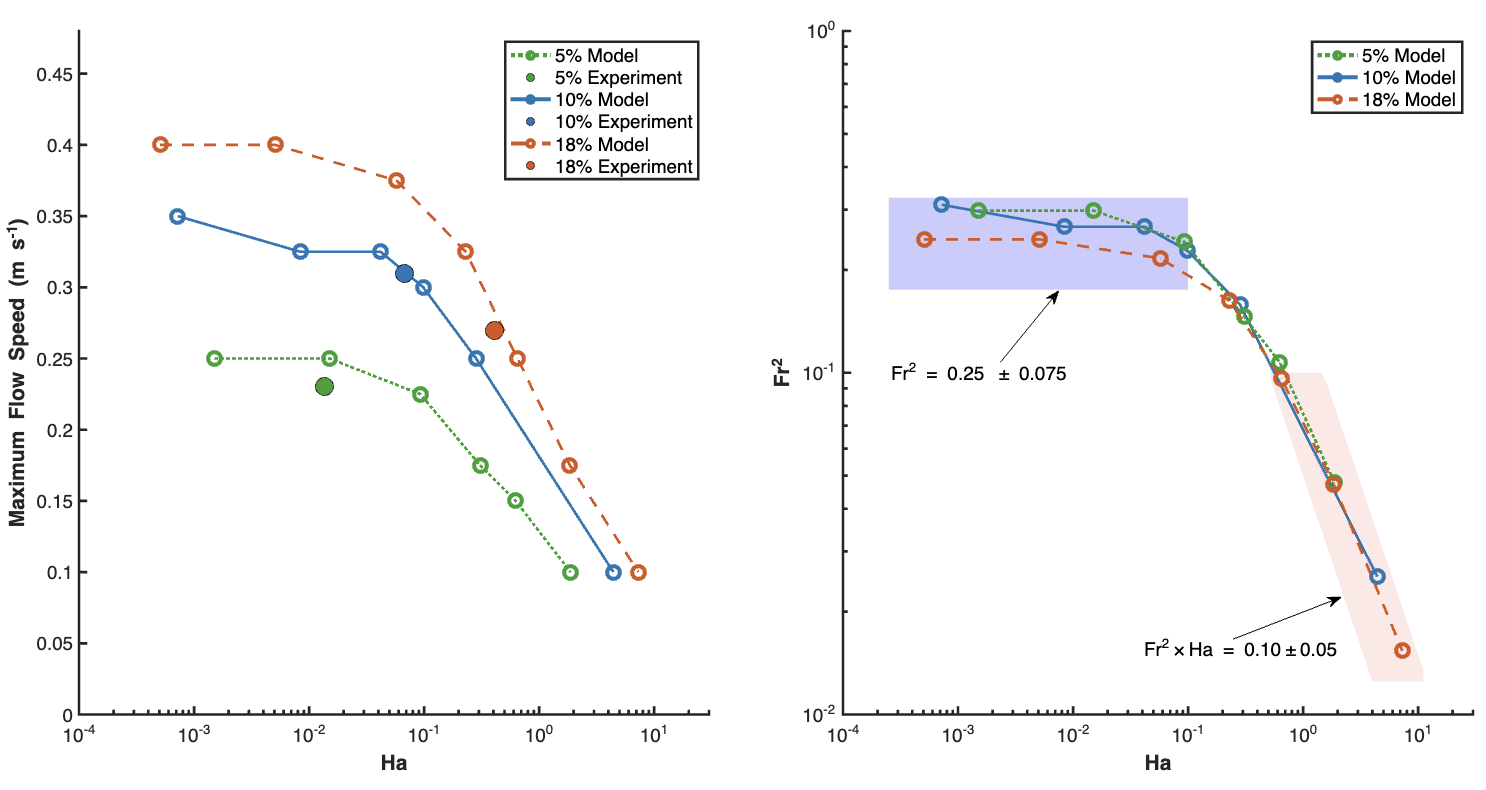}
    \caption{(\textbf{Left}) Maximum flow speed as a function of Hampton number \textit{Ha} for cohesive SGFs shown in Fig. \ref{fig:YieldStressInfluence} with initial bentonite fractions of 5\%, 10\%, and 18\%. Model results (open circles, with dashed lines to guide the eye) and experimental data (filled circles) exhibit two distinct regimes separated by a change in slope: a low \textit{Ha} regime (turbidity currents) characterized by weak sensitivity of maximum flow speed to \textit{Ha}, and a high \textit{Ha} regime (debris flows) where maximum flow speed decreases sharply with increasing \textit{Ha}. The transition between regimes occurs near \textit{Ha} $\approx$ $10^{-2}$ with consistent behavior across fractions. (\textbf{Right}) Densimetric Froude Number \textit{Fr} (plotted as $Fr^2$ to reflect a stress ratio)  as a function of \textit{Ha} for the same SGFs. The \textit{Fr} number exhibits a similar two-regime structure: a low \textit{Ha} plateau where inertial and buoyancy forces remain in near-constant proportion, and a sharp decline at high \textit{Ha} consistent with the transition to a debris flow regime.}
    \label{fig:Hampton}
\end{figure}

In this last section, we present a series of simulations where we systematically alter sediment solid fraction $\phi_s$ and sediment yield stress $\tau$ over a wide range of values in order to help elucidate fundamental controls on the flow. In the solver used in previous sections, $\tau$ is a function of $\phi_s$ through Eq. \ref{Q1}, but here the two variables are de-coupled to highlight the competing roles of solid fraction and yield stress in governing cohesive SGF behavior. Results are presented as a qualitative regime map (Fig. \ref{fig:YieldStressInfluence}) that visualizes the influence of sediment fraction and yield stress on flow morphology. Across the range considered, increasing solid content generally produces more coherent, compact, and mobile currents, with clearer head formation and longer front propagation, reflecting the stronger density contrast and reduced tendency for early dilution. In contrast, increasing yield stress consistently suppresses mobility, shortens runout, and promotes thicker, more localized deposits, indicating that viscous-plastic resistance increasingly dominates over gravitational driving forces. The intermediate cases present the most dynamically complex behavior, where sufficiently mobile fronts coexist with a more resistant trailing body, producing distinct head–tail structure and interfacial instability. Based on the results presented in Fig. \ref{fig:YieldStressInfluence}, we propose that SGFs can be classified using two non-dimensional numbers. The first number, shown as the horizontal axis in Fig. \ref{fig:Hampton}, is referred to here as the Hampton number \textit{Ha} (Eq. \ref{Equation:Hampton}) \cite{Du2022,Choi2024}. It reflects the ratio of sediment yield stress to fluid inertial stress and explains the non-monotonic relation between SGF speed and sediment fraction shown in Fig \ref{fig:ComparingRunout}.
\begin{equation}
Ha=\frac{\tau}{\rho_{b} u^2}
  \label{Equation:Hampton}
\end{equation}

In Eq. \ref{Equation:Hampton}, $\rho_{b}$ is the initial bulk density of the sediment bed, $\tau$ is the yield stress, and $u$ is the maximum SGF flow speed. The second number shown as the vertical axis in Fig. \ref{fig:Hampton}, is a densimetric Froude number characterizing the ratio of fluid inertial stress to gravity driving force \cite{Mohrig1998}.
\begin{equation}
Fr^2=\frac{\rho_bu^2}{\sqrt{(\rho_{b}-\rho_{w})gh}}
  \label{Equation:DensFroude}
\end{equation}

In Eq. \ref{Equation:DensFroude}, $\rho_{w}$ is the density of water, \textit{g} is gravitational acceleration, \textit{h} is initial sediment height. We note that \textit{Fr} is more commonly expressed using fluid density $\rho_w$ in the inertial term, such that $Fr=\frac{u}{\sqrt{\left(\frac{\rho_b}{\rho_w}-1\right)gh}}$. Here we used $\rho_b$ in the inertial term for consistency with Eq. \ref{Equation:Hampton}. As seen in Fig. \ref{fig:Hampton}, the \textit{Ha} number classifies SGFs into two primary regimes. The first regime, turbidity currents, has a maximum SGF speed determined by the balance of gravity and inertial phenomena. This regime is expressed as a quasi-plateau relationship between \textit{Ha} and \textit{Fr} (or equivalently, between \textit{Ha} and \textit{U}) in Fig. \ref{fig:Hampton}. 

The second regime corresponds to the debris flow regime, in which flow mobility is primarily controlled by \textit{Ha}, and front speed decreases as yield stress effects become more important. Identifying this regime boundary is useful because it helps determine whether a small increase in yield stress or bulk density will exert the stronger control on clay-rich flow mobility. A similar sensitivity has been observed in sand–clay mixtures, in which small additions of sand may increase or decrease mobility depending on SGF regime \cite{Baker2023}.

Finally, the product of \textit{Fr} and \textit{Ha} determines whether gravitational effects are sufficiently large to overcome sediment yield stress. In other words, flow initiation requires that $Fr^2$ $\times$ \textit{Ha} = $\tau$/[$\left(\rho_b-\rho_w\right)gh$] $<$$<$ 1. This is consistent with the observations that the relationship between $Fr^2$ and \textit{Ha} has a slope of -1 at high Ha values on the log-log representation in Fig. \ref{fig:Hampton}. In this high-\textit{Ha} regime, the value of $Fr^2\times Ha$ is relatively uniform and on the order of 0.1.

Taken together, these results suggest that cohesive SGF behavior and regime type is controlled by a balance between buoyancy-driven propagation via flow bulk densities and yield-stress arrest resulting from higher rheological strength, with transitional regimes emerging where neither mechanism fully dominates. This solid fraction–yield stress framework provides a basis for interpreting the observed variability in front speed, flow and deposit morphology, and runout across the broader parameter space explored in this study. Combined with geometric considerations, this demonstrates the importance of the potential energy of the initial configuration and of the non-Newtonian rheology of cohesive sediments. 
\section{Conclusion}
This study demonstrates that cohesive sediment gravity flow behavior can be accurately predicted using a CFD framework that incorporates realistic rheological descriptions and interphase momentum transfer. Model results show strong agreement with experimental observations from \citeA{Baker2017}, successfully reproducing front morphology, maximum flow speeds, and flow regimes across two different clay mineralogies and a wide range of initial sediment solid fractions. In addition, runout behavior trends are correctly predicted, but runout distance values are underestimated. 

Across all simulations, sediment gravity flow mobility is primarily governed by the interplay between yield stress–controlled resistance to sediment deformation, inertial resistance to bulk fluid displacement, and density difference driving forces. Increasing yield stress systematically reduces flow velocity and promotes earlier flow arrest, while increasing sediment bulk density, if decoupled from yield stress, has the opposite effect. These competing mechanisms define distinct flow regimes and explain observed transitions between low-density turbidity currents, high-density turbidity currents, mudflows, and mudslides.

Parametric analyses reveal that both intrinsic (e.g., rheology, sediment density) and extrinsic (sediment fraction and initial sediment height) properties exert strong controls on maximum flow velocity and overall flow evolution. In particular, yield stress is shown to systematically influence flow mobility, with regime transitions and flow characteristics well described by non-dimensional metrics such as the Hampton and Froude numbers. 

While the model captures key aspects of flow dynamics, including front propagation and bulk mobility, some discrepancies remain in the detailed deposit morphology and runout extent, particularly at lower sediment fractions. These differences likely reflect limitations in representing thixotropic effects and particle settling, particularly at low solid fractions. Future work should explore the role of thixotropic rheological behavior in controlling late-stage flow dynamics and deposition.

Overall, the results presented here provide a mechanistic basis for predicting erosion, transport, and deposition of cohesive sediment gravity flows across marine, lacustrine, and reservoir environments based on laboratory measurements of sediment rheology, and in particular, measurements of sediment yield stress as a function of solid fraction (Fig. \ref{fig:YS_Fit}). More broadly, this work suggests the capability of CFD-based approaches to help bridge laboratory and field scale observations. Future extensions incorporating granular sediment dynamics through coupling with discrete element method (DEM) approaches and biogeochemical effects (through inclusion of the rheological impacts of organic matter and EPS) offer a promising pathway toward more comprehensive and predictive models of cohesive sediment transport \cite{Vowinckel2023,Qin2024}. Future extensions also have the potential to contribute to our understanding of near-shore sediment transport during extreme events \cite{Sherwood2022} and to terrestrial geophysical hazards including mudslides and debris flows \cite{Iverson1997,Pudasaini2012}.

\section*{Open Research Section}
Data is available in a preview of Zenodo Database
\section*{Conflict of Interest disclosure}
The authors declare there are no conflicts of interest for this manuscript.

\acknowledgments

This research was supported by the US DOE Office Of Science, Geosciences Program (Award DE-SC0018419). MDJ was additionally supported by the High Meadows Environmental Institute at Princeton University through the Mary and Randall Hack ‘69 Fund for Water Research. JQY was supported by NSF Award EAR-2150796. CS was supported by the French National Research Agency (ANR) under grant number ANR-22-CE30-0017, project PhysErosion.
Simulations were performed using computational resources managed and supported by Princeton Research Computing, a consortium of groups including the Princeton Institute for Computational Science and Engineering (PICSciE) and the Office of Information Technology's High Performance Computing Center and Visualization Laboratory at Princeton University. Any opinions, findings, and conclusions or recommendations expressed in this material are those of the authors and do not necessarily reflect the views of the Department of Energy and National Science Foundation.

%
%

\bibliography{Bibliography}

@article{Jeong2010,
   author = {Sueng Won Jeong and Jacques Locat and Serge Leroueil and Jean Philippe Malet},
   doi = {10.1139/T10-012},
   issn = {00083674},
   issue = {10},
   journal = {Canadian Geotechnical Journal},
   pages = {1085-1100},
   title = {Rheological properties of fine-grained sediment: The roles of texture and mineralogy},
   volume = {47},
   year = {2010}
}

@article{Jeong2012,
   author = {Sueng Won Jeong and Jacques Locat and Serge Leroueil},
   doi = {10.1346/CCMN.2012.0600202},
   issn = {00098604},
   issue = {2},
   journal = {Clays and Clay Minerals},
   month = {4},
   pages = {108-120},
   title = {The effects of salinity and shear history on the rheological characteristics of illite-rich and na-montmorillonite-rich clays},
   volume = {60},
   year = {2012}
}

@article{Yang2014,
   author = {Wen-Yu Yang and Guo-Liang Yu and Soon Keat Tan and Hua-Kun Wang},
   issue = {4},
   journal = {International Journal of Sediment Research},
   pages = {454-470},
   title = {Rheological properties of dense natural cohesive sediments subject to shear loadings},
   volume = {29},
   year = {2014},
   doi = {https://doi.org/10.1016/S1001-6279(14)60059-7}
}

@article{Chauchat2017,
   author = {Julien Chauchat and Zhen Cheng and Tim Nagel and Cyrille Bonamy and Tian Jian Hsu},
   doi = {10.5194/gmd-10-4367-2017},
   issn = {19919603},
   issue = {12},
   journal = {Geoscientific Model Development},
   month = {11},
   pages = {4367-4392},
   publisher = {Copernicus GmbH},
   title = {SedFoam-2.0: A 3-D two-phase flow numerical model for sediment transport},
   volume = {10},
   year = {2017}
}

@article{Qin2024,
   author = {Chenxi Qin and Lunliang Duan and Duoyin Wang and Bingchuan Duan and Wei Liu},
   doi = {10.1063/5.0207743},
   issn = {10897666},
   issue = {5},
   journal = {Physics of Fluids},
   month = {5},
   publisher = {American Institute of Physics},
   title = {A local scour model for single pile on silty seabed considering soil cohesion (SedCohFOAM): Model and validation},
   volume = {36},
   year = {2024}
}

@article{Kurz2022,
   author = {Dorothee L Kurz and Eleonora Secchi and Francisco J Carrillo and Ian C Bourg and Roman Stocker and Joaquin Jimenez-Martinez},
   doi = {10.1073/pnas},
   journal = {Proceedings of the National Academy of Sciences},
   title = {Competition between growth and shear stress drives intermittency in preferential flow paths in porous medium biofilms},
   year = {2022},
   pages = {1-10},
   volume = {119}
}

@book{Mehta2022,
   author = {Ashish Mehta},
   edition = {2},
   publisher = {World Scientific Publishing },
   title = {An Introduction to Hydraulics of Fine Sediment Transport },
   doi ={https://doi.org/10.1142/12873},
   year = {2022}
}

@article{Neale1974,
   author = {Graham Neale and Walter Nader},
   doi = {10.1002/cjce.5450520407},
   issn = {1939019X},
   issue = {4},
   journal = {The Canadian Journal of Chemical Engineering},
   pages = {475-478},
   title = {Practical significance of brinkman's extension of darcy's law: Coupled parallel flows within a channel and a bounding porous medium},
   volume = {52},
   year = {1974}
}

@article{Schleiss2016,
   author = {Anton J. Schleiss and Mário J. Franca and Carmelo Juez and Giovanni De Cesare},
   doi = {10.1080/00221686.2016.1225320},
   issn = {00221686},
   issue = {6},
   journal = {Journal of Hydraulic Research},
   pages = {595-614},
   publisher = {Taylor and Francis Ltd.},
   title = {Reservoir sedimentation},
   volume = {54},
   year = {2016},
   doi = {https://doi.org/10.1080/00221686.2016.1225320}
}

@article{Sposito1999,
   author = {Garrison Sposito and Neal T Skipper and Rebecca Sutton and Sung-Ho Park and Alan K Soper and Jeffery A Greathouse},
   pages = {3358-3364},
   title = {Surface geochemistry of the clay minerals},
   volume = {96},
   year = {1999},
   journal = {Proceedings of the National Academy of Science},
   doi = {https://doi.org/10.1073/pnas.96.7.335}
}

@article{Carrillo2019,
   author = {Francisco J. Carrillo and Ian C. Bourg},
   doi = {10.1029/2019WR024712},
   issn = {19447973},
   issue = {10},
   journal = {Water Resources Research},
   month = {10},
   pages = {8096-8121},
   publisher = {Blackwell Publishing Ltd},
   title = {A Darcy-Brinkman-Biot Approach to Modeling the Hydrology and Mechanics of Porous Media Containing Macropores and Deformable Microporous Regions},
   volume = {55},
   year = {2019}
}

@article{Carrillo2021,
   author = {Francisco J. Carrillo and Ian C. Bourg},
   doi = {10.1029/2020WR028734},
   issn = {19447973},
   issue = {2},
   journal = {Water Resources Research},
   month = {2},
   publisher = {Blackwell Publishing Ltd},
   title = {Modeling Multiphase Flow Within and Around Deformable Porous Materials: A Darcy-Brinkman-Biot Approach},
   volume = {57},
   year = {2021},
   pages = {1-27}
}

@article{Baker2017,
   author = {M L Baker and J H Baas and R S Jacinto and M J Craig and I A Kane and S Barker},
   journal = {Journal of Sedimentary Research},
   pages = {1176-1195},
   title = {The Effect of Clay Type On the Properties of Cohesive Sediment Gravity Flows and Their Deposits},
   year = {2017},
   doi = {https://doi.org/10.2110/jsr.2017.63},
   volume = {87}
}

@article{Locat2002,
   author = {Jacques Locat and Homa J Lee},
   journal = {Canadian Geotechnical Journal},
   pages = {193-212},
   title = {Submarine Landslides: Advances and Challenges},
   year = {2002},
   doi = {https://doi.org/10.1139/t01-089},
   volume = {39}
}

@article{Galy2007,
   author = {Valier Galy and Christian France-Lanord and Olivier Beyssac and Pierre Faure and Hermann Kudrass and Fabien Palhol},
   journal = {Nature},
   pages = {407-410},
   title = {Efficient Organic Carbon Burial in the Bengal Fan Sustained by the Himalayan Erosional System},
   year = {2007},
   doi = {https://doi.org/10.1038/nature06273},
   volume = {450}
}

@article{Marr2001,
   author = {Jeffrey G Marr and Peter A Harff and G Shanmugam and Gary Parker},
   journal = {Geological Society of America Bulletin},
   pages = {1377-1386},
   title = {Experiments on Subaqueous sandy gravel flows: the role of clay and water content in flow dynamics and depositional structures},
   year = {2001},
   doi = {https://doi.org/10.1130/0016-7606(2001)113<1377:EOSSGF>2.0.CO;2},
   volume = {113}
}

@article{Bottner2024,
   author = {Christoph Bottner and Christopher J Stevenson and Rebecca Englert and Mischa Schonke and Bruna T Pandolpho and Jacob Geerson and Peter Feldens and Sabastian Krastel},
   journal = {Science Advances},
   title = {Extreme Erosion and Bulking in a Giant Submarine Gravity Flow},
   year = {2024},
   doi = {DOI: 10.1126/sciadv.adp258},
   volume ={10}
}

@article{Talling2015,
   author = {Peter J Talling and Joshua Allin and Dominic A Armitage and Robert W C Arnott and Matthieu JB Cartigny and Michael A Clare and Fabrizio Felletti and Jacob A Covault and Stephanie Giradclos and Ernst Hansen and Philip R Hill and Richard N Hiscott and Andrew J Hogg and John Hughes Clarke and Zane R Jobe and Giuseppe Malgesini and Alessandro Mozzato and Hajime Naruse and Sam Parkinson and Frank J peel and David JW Piper and Ed Pope and George Postma and Pete Rowley and Andrew Sguazzini and Christopher J Stevenson and Esther J Sumner and Zoltan Sylvester and Camilla Watts and Jinping Xu},
   journal = {Journal of Sedimentary Research},
pages = {153-169},
   title = {Key Future Directions for Research on Turbidity Currents and Their Deposits},
   year = {2015},
   doi = {https://doi.org/10.2110/jsr.2015.03},
   volume = {85}
}

@article{Carrillo2021b,
   author = {Francisco J. Carrillo and Ian C. Bourg},
   doi = {10.1103/PhysRevE.103.063106},
   issn = {24700053},
   issue = {6},
   journal = {Physical Review E},
   month = {6},
   pmid = {34271761},
   publisher = {American Physical Society},
   title = {Capillary and viscous fracturing during drainage in porous media},
   volume = {103},
   year = {2021},
   pages= {1-10}
}

@article{Guo2024,
   author = {Xingsen Guo and Xiaolei Liu and Tianyuan Zhenga and Hong Zhange and Yang Lu and Tiantao Li},
   journal = {Coastal Engineering},
   pages = {1-14},
   title = {A mass transfer-based LES modelling methodology for analyzing the
movement of submarine sediment flows with extensive shear behavior},
    year = {2024},
    doi = {https://doi.org/10.1016/j.coastaleng.2024.104531},
    volume = {191}
}

@article{Stevenson2018,
   author = {Christopher John Stevenson and  Peter Feldens and Aggeliki Georgiopoulou and Mischa Schonke and Sebastian Krastel and David J.W. Piper and Katja Lindhorst and David Mosher },
   journal = {Nature Communications},
   pages = {1-7},
   title = {Reconstructing the sediment concentration of a
giant submarine gravity flow},
    year = {2018},
    doi = {https://doi.org/10.1038/s41467-018-05042-6},
    volume = {9}
}

@article{Du2022,
   author = {Jianting Du and  Clarence Edward Choi and Jiantao Yu and Vikas Thakur},
   journal = {Journal of
Geophysical Research: Earth Surface},
   pages = {1-23},
   title = {Mechanisms of Submarine Debris Flow Growth},
    year = {2022},
    doi = {https://doi.org/10.1029/2021JF006470},
    volume ={127}
}

@article{Zhou2025,
   author = {Shu Zhou and  Xiaolin Tan and Jian Pu and Zhen Guo and Chongqiang Zhu and Jin Sun and Yu Huan},
   journal = {Engineering Geology},
   pages = {1-15},
   title = {The Effects of Clay Content on the Dynamics of Submarine Landslides: New Insights from Flume Experiments},
    year = {2025},
    doi = {https://doi.org/10.1016/j.enggeo.2025.108157},
    volume={353}
}

@article{Craig2019,
   author = {Melissa J Craig and Jaco H Baas and Kathryn J Amos and Lorna J Strachan and Andrew J Manning and David M Paterson and Julie A Hope and Scott D Nodder and Megan L Baker},
   journal = {Geology},
   pages = {72-76},
   title = {Biomediation of Submarine Sediment Gravity Flow Dynamics},
    year = {2019},
    doi = {https://doi.org/10.1130/G46837.1},
    volume = {48}
}

@article{Meiburg2010,
   author = {Eckart Meiburg and Ben Kneller},
   journal = {Annual review of fluid mechanics},
   pages = {135-156},
   title = {Turbidity currents and their deposits},
    year = {2010},
    doi = {https://doi.org/10.1146/annurev-fluid-121108-145618},
    volume ={42}
}

@article{Jerolmack2019,
   author = {Doug J Jerolmack and Karen E Daniels},
   journal = {Nature Reviews Physics},
   pages = {716-730},
   title = {Viewing Earth’s surface as a soft-matter landscape},
    year = {2019},
    doi = {https://doi.org/10.1038/s42254-019-0111-x},
    volume = {1}
}

@article{Dasgupta2003,
   author = {Prabir Dasgupta},
   journal = {Earth-science reviews},
   pages = {265-281},
   title = {Sediment gravity flow-the conceptual problems},
    year = {2003},
    doi = {https://doi.org/10.1016/S0012-8252(02)00160-5},
    volume ={62}
}

@article{Pohl2020,
   author = {Florian Pohl and Joris T Eggenhuisen and Ian A Kane and Michael A Clare},
   journal = { Environmental science \& technology},
   pages = {4180-4189},
   title = {Transport and Burial of Microplastics in Deep-Marine Sediments by
Turbidity Currents},
    year = {2020},
    doi = {https://doi.org/10.1021/acs.est.9b07527},
    volume = {54}
}

@article{Mesri1971,
   author = {Gholamreza Mesri and Roy E Olson},
   journal = {Clays and Clay mineral},
   pages = {151-158},
   title = {Mechanisms controlling the permeability of clays},
    year = {1971},
    doi = {https://doi.org/10.1346/CCMN.1971.0190303},
    volume = {19}
}

@article{Haza2013,
   author = {Zainul Faizien Haza and Indra Sati Hamonangan Harahap and Lema Mosissa Dakssa},
   journal = {Natural Hazards},
   pages = {587-611},
   title = {Experimental studies of the flow-front and drag forces exerted by subaqueous mudflow on inclined base},
    year = {2013},
    doi = {https://doi.org/10.1007/s11069-013-0643-9},
    volume = {68}
}

@article{Sobocinska2022,
   author = {Alicja Sobocinska and Jaco H Baas},
   journal = {Marine Geology},
   title = {Effect of biological polymers on mobility and run-out distance of cohesive and non-cohesive sediment gravity flows},
    year = {2022},
    doi = {https://doi.org/10.1016/j.margeo.2022.106904},
    pages ={1-13},
    volume = {452}
}

@book{Velde1995,
    editor = {Bruce Velde},
    title = {Origin and Mineralogy of Clays} ,
    publisher ={Springer},
    year = {1995},
    doi ={https://doi.org/10.1007/978-3-662-12648-6}
}

@article{Ito2017,
   author = {Akihiko Ito and Rota Wagai},
   journal = {Nature: Scientific Data},
   pages = {1-11},
   title = {Global distribution of clay-size minerals on land surface for biogeochemical and climatological studies},
    year = {2017}
}

@article{Hesse2006,
   author = {Reinhard Hesse and Saeed Khodabakhsh},
   journal = {Sedimentary Geology},
   pages = {1-11},
   title = {Significance of fine-grained sediment lofting from melt-water generated turbidity currents for the timing of glaciomarine sediment
transport into the deep sea},
    year = {2006},
    doi = {https://doi.org/10.1016/j.sedgeo.2005.10.006},
    volume = {186}
}

@article{Mohrig1998,
   author = {David Mohrig and Chris Ellis and Gary Parker and Kelin X Whipple and Midhat Hondzo},
   journal = {Geological Society of America Bulletin},
   pages = {387-394},
   title = {Hydroplaning of subaqueous debris flows},
    year = {1998},
    doi = {https://doi.org/10.1130/0016-7606(1998)110<0387:HOSDF>2.3.CO;2},
    volume = {110}
}

@article{Whorton2025,
   author = {Jodie Whorton and Thomas J Jones and Lionel Wilson},
   journal = {Earth Science Reviews},
   title = {Mudflow rheology: A review and analysis for earth and planetary science disciplines},
    year = {2025},
    doi = {https://doi.org/10.1016/j.earscirev.2025.105226},
    pages= {1-23}
}

@article{Coussot1995,
   author = {Pierre Coussot},
   journal = {Physical Review Letters},
   pages = {3971-3974},
   title = {Structural similarity and transition from newtonian to non-newtonian behavior for clay-water suspensions},
    year = {1995},
    doi = {https://doi.org/10.1103/PhysRevLett.74.3971},
    volume = {74}
}

@article{Chamoun2016,
   author = {Sabine Chamoun and Giovanni De Cesare and Anton J Schleiss },
   journal = {International Journal of Sediment Research},
   pages = {195-204},
   title = {Managing reservior sedimentation by venting turbidity currents: a review},
    year = {2016},
    doi = {https://doi.org/10.1016/j.ijsrc.2016.06.001},
    volume ={31}
}

@article{Felix2006,
   author = {Maarten Felix and Jeffrey Peakalln},
   title = {Transformation of debris flows into turbidity currents:
mechanisms inferred from laboratory},
 journal = {Sedimentology},
pages = {107-123},
    year = {2006},
    doi = {https://doi.org/10.1111/j.1365-3091.2005.00757.x},
    volume = {53}
}

@article{Zheng2025,
   author = {Xiaojin Zheng and Xinyi Shen and Ian C Bourg},
   title = {Coarse-grained simulation of colloidal self-assembly, cation exchange, and rheology in Na/Ca smectite clay gels},
 journal = {Journal of Colloid and Interface Science},
    year = {2025},
    doi = {https://doi.org/10.1016/j.jcis.2025.137573},
    pages = {1-12},
    volume = {693}
}

@article{VonBoetticher2016,
   author = {Albrecht von Boetticher and Jens M Turowski and Brian W McArdell and Dieter Rickenmann and James W Kirchner},
   title = {DebrisInterMixing-2.3: A finite volume solver for three-dimensional debris-flow simulations with two calibration parameters- Part 1: Model description},
pages = {2909-2923},
 journal = {Geoscientific Model Development},
    year = {2016},
    doi = {https://doi.org/10.5194/gmd-9-2909-2016},
    volume = {9}
}

@article{Choi2024,
   author = {Clarence Edward Choi and Jiantao Yu and Jiaqi Zhang},
   title = {Review of the missing link between field and modeled submarine debris flows: Scale effects of physical modeling},
 journal = {Earth-Science Reviews },
    year = {2024},
    doi = {https://doi.org/10.1016/j.earscirev.2024.104911a},
    pages = {1-17},
    volume={258}
}

@article{Zadehali2025,
   author = {Ehsan Zadehali and Soukaina Benaich and Shih-Hsun Huang and Ian Bourg and Judy Yang},
   title = {Salinity reduces yield stress and erosion threshold in sand‐clay mixtures: Evidence from rheometry and flume experiments},
 journal = {Water Resources Research},
year = {2025},
doi = {https://doi.org/10.1029/2024WR039529},
pages = {1-14},
volume = {61}
}

@article{He2018,
   author = {Zhiguo He and Liang Zhao and Peng Hu and Chinghao Yu and Ying-Yien Lin},
   title = {Investigations of dynamic behaviors of lock-exchange turbidity currents down a slope based on direct numerical simulation},
 journal = {Advances in Water Resources},
year = {2018},
doi = {https://doi.org/10.1016/j.advwatres.2018.07.008},
pages = {164-177},
volume = {119}
}

@article{Issa1986,
   author = {Raad I Issa},
   title = {Solution of the implicitly discretised fluid flow equations by operator-splitting},
pages = {40-65},
 journal = {Journal of computational physics},
year = {1986},
doi = {https://doi.org/10.1016/0021-9991(86)90099-9},
volume = {62}
}

@article{Ancey2007,
   author = {Christophe Ancey},
   title = {Plasticity and geophysical flows: A review},
pages = {4-35},
 journal = {Journal of non-Newtonian fluid mechanics},
year = {2007},
doi = {https://doi.org/10.1016/j.jnnfm.2006.05.005},
volume = {142}
}

@article{Enwald1996,
   author = {Hans Enwald and Eric Peirano and A-E Almstedt},
   title = {Eulerian two-phase flow theory applied to fluidization},
pages = {21-66},
 journal = {International Journal of Multiphase Flow},
year = {1996},
doi = {https://doi.org/10.1016/S0301-9322(96)90004-X},
volume = {22}
}

@article{Mohrig2003,
   author = {David Mohrig and Jeffrey G Marr},
   title = {Constraining the efficiency of turbidity current generation from submarine debris flows and slides using laboratory experiments},
pages = {883-899},
 journal = {Marine and Petroleum Geology},
year = {2003},
doi ={https://doi.org/10.1016/j.marpetgeo.2003.03.002},
volume = {20}
}

@article{Talling2023,
   author = {Peter J Talling and Matthieu JB Catigny and Ed Pope and Megan Baker and Michael A Clare and Maarten Heijnen and Sophie Hage and Dan R Parsons and Steve M Simmons and Charlie K Pauli and Roberto Gwiazda and Gwyn Lintern and John E Hughes Clarke and Jingping Xu and Ricardo Silva Jacinto and Katherine L Maier},
   title = {Detailed monitoring reveals the nature of submarine turbidity currents},
pages = {642-658},
 journal = {Nature Reviews Earth \& Environment},
year = {2023},
doi = {https://doi.org/10.1038/s43017-023-00458-1},
volume = {4}
}

@article{Iverson1997,
   author = {Richard M Iverson},
   title = {The physics of debris flows},
pages = {245-296},
 journal = {Reviews of geophysics},
year = {1997},
doi = {https://doi.org/10.1146/annurev.earth.25.1.85},
volume = {25}
}

@article{Hampton1996,
   author = {Monty A Hampton and Homa J Lee and Jacques Locat},
   title = {Submarine landslides},
pages = {33-59},
 journal = {Reviews of geophysics},
year = {1996},
doi ={https://doi.org/10.1029/95RG03287},
volume = {34}
}

@article{Middleton1973,
   author = {Gerard V Middleton and Monty A Hampton},
   title = {Part I. Sediment gravity flows: mechanics of flow and deposition},
pages = {1-38},
year = {1973}
}

@article{Peng2025,
   author = {Chen Peng and Xingyue Li and Yu Huang},
   title = {Dynamics and mechanisms of subaqueous gravity flows composed of viscous liquids and particles},
 journal = {Bulletin of Engineering Geology and the Environment},
year = {2025},
doi = {https://doi.org/10.1007/s10064-025-04465-y},
pages = {1-19},
volume= {84}
}

@article{Rzadkiewicz1997,
   author = {Assier S Rzadkiewicz and C Mariotti and Philippe Heinrich},
   title = {Numerical simulation of submarine landslides and their hydraulic effects},
pages = {149-157},
 journal = {Journal of Waterway, Port, Coastal, and Ocean Engineering},
year = {1997},
doi = {https://doi.org/10.1061/(ASCE)0733-950X(1997)123:4(149)},
volume = {123}
}

@article{Li2024,
   author = {Xiaohan Li and Ian C Bourg},
   title = {Hygroscopic growth of adsorbed water films on smectite clay particles},
pages = {1109-1118},
 journal = {Environmental Science \& Technology},
year = {2024},
doi = {https://doi.org/10.1021/acs.est.3c08253},
volume = {2024}
}

@article{Talling2013,
   author = {Peter J Talling},
   title = {Hybrid submarine flows comprising turbidity current and cohesive debris flow: Deposits, theoretical and experimental analyses, and generalized model},
pages = {460-488},
 journal = {Geosphere},
year = {2013},
doi = {https://doi.org/10.1130/GES00793.1},
volume = {9}
}

@article{Baas2002,
   author = {Jaco H Baas and James L Best},
   title = {Turbulence modulation in clay-rich sediment-laden flows and some implications for sediment deposition},
pages = {336-340},
 journal = {Journal of Sedimentary Research},
year = {2002},
doi = {https://doi.org/10.1306/120601720336},
volume ={72}
}

@article{Jing2018,
   author = {Lu Jing and C Y Kwok and Y F Leung and Zirui Zhang and L Dai},
   title = {Runout scaling and deposit morphology of rapid mudflows},
pages = {2004-2023},
 journal = {Journal of Geophysical Research: Earth Surface},
year = {2018},
doi ={ https://doi.org/10.1029/2018JF004667},
volume = {123}
}

@inbook{Guven1992,
    author = {Necip Güven},
    title = {Molecular Aspects of Clay-Water Interactions},
    booktitle ={Clay-Water Interface and its Rheological Implications} ,
    year = {1992},
    chapter = {1}
}

@article{Hermidas2018,
  author  = {Navid Hermidas and Joris T. Eggenhuisen and Ricardo Silva Jacinto and Stefan M. Luthi and Ferenc Toth and Florian Pohl},
  title   = {A classification of clay-rich subaqueous density flow structures},
  pages   = {945--966},
  journal = {Journal of Geophysical Research: Earth Surface},
  year    = {2018},
  doi = { https://doi.org/10.1002/2017JF004386},
  volume = {123}
}

@article{Baas2009,
   author = {Jaco H Baas and James L Best and Jeffrey Peakall and M I Wang},
   title = {A phase diagram for turbulent, transitional, and laminar clay suspension flows},
pages = {162-183},
 journal = {Journal of Sedimentary Research},
year = {2009},
doi = {https://doi.org/10.2110/jsr.2009.025},
volume = {79}
}

@article{Mead2017,
   author = {Stuart R Mead and Christina Magill and Vincent Lemiale and Jean-Claude Thouret and Makesh Prakash},
   title = {Examining the impact of lahars on buildings using numerical modelling},
pages = {703-719},
 journal = {Natural Hazards and Earth System Sciences},
year = {2017},
doi = {https://doi.org/10.5194/nhess-17-703-2017},
volume = {17}
}

@article{Giri2006,
   author = {Sanjay Giri and Yasuyuki Shimizu},
   title = {Numerical computation of sand dune migration with free surface flow},
 journal = {Water Resources Research},
year = {2006}, 
doi ={https://doi.org/10.1029/2005WR004588},
pages = {1-19},
volume = {42}
}

@article{Oda2011,
   author = {Kenichi Oda and Shuji Moriguchi and Isao Kamiishi and Atsushi Yashima and Kazuhide Sawarda and Atsushi Sato},
   title = {Simulation of a snow avalanche model test using computational fluid dynamics},
pages = {57-64},
 journal = {Annals of Glaciology},
year = {2011},
doi = {https://doi.org/10.3189/172756411797252284},
volume = {52}
}

@article{Baker2023,
   author = {Megan L Baker and Jaco H Baas},
   title = {Does sand promote or hinder the mobility of cohesive sediment gravity flows?},
pages = {1110-1130},
 journal = {Sedimentology},
year = {2023},
doi = {https://doi.org/10.1111/sed.13072},
volume ={70}
}

@article{vanKessel1998,
   author = {Thijs Van Kessel and C Blom},
   title = {Rheology of cohesive sediments: comparison between a natural and an artificial mud},
pages = {591-612},
 journal = {Journal of Hydraulic Research},
year = {1998},
doi = {https://doi.org/10.1080/00221689809498611},
volume = {36}
}

@article{Spearman2017,
   author = {Jeremy Spearman},
   title = {An examination of the rheology of flocculated clay suspensions},
pages = {485-497},
 journal = {Ocean Dynamics},
year = {2017},
doi = {https://doi.org/10.1007/s10236-017-1041-8},
volume = {67}
}

@article{Zheng2025B,
   author = {Xiaojin Zheng and Ian C Bourg},
   title = {A multiscale approach to simulate non‐isothermal multiphase flow in deformable porous materials},
 journal = {Water Resources Research },
year = {2025},
doi = {https://doi.org/10.1029/2025WR041300},
pages ={1-21},
volume = {61}
}

@article{Jeon2018,
   author = {Chan-Hoo Jeon and Ben R. Hodges},
   title = {Comparing thixotropic and Herschel–Bulkley parameterizations for continuum models of avalanches and subaqueous debris flows},
 journal = {Natural Hazards and Earth System Sciences},
year = {2018},
doi = {https://doi.org/10.5194/nhess-18-303-2018},
pages= {303-319},
volume = {18}
}

@article{Trujillo-Vela2022,
   author = {Mario Germán Trujillo-Vela and Alfonso Mariano Ramos-Cañón and Jorge Alberto Escobar-Vargas and Sergio Andres Galindo Torres},
   title = {An overview of debris-flow mathematical modelling},
 journal = {Earth-Science Reviews},
year = {2022},
doi = {https://doi.org/10.1016/j.earscirev.2022.104135s},
pages = {1-28},
volume = {232}
}

@article{Pignon1997,
   author = {Frédéric Pignon and Albert Magnin and Jean-Michel Piau and Bernard Cabane and Peter Lindner and Olivier Diat},
   title = {Yield stress thixotropic clay suspension: Investigations of structure by light, neutron, and x-ray scattering},
 journal = {Physical Review E},
year = {1997},
doi ={10.1103/PhysRevE.56.3281},
pages = {3281-3289},
volume = {56}
}

@article{Soulaine2015,
   author = {Cyprien Soulaine and Michel Quintard and Hervé Allain and Bertrand Baudouy and Rob Van Weelderen},
   title = {A PISO-like algorithm to simulate superfluid helium flow with the two-fluid model},
 journal = {Computer Physics Communications},
year = {2015},
doi = {https://doi.org/10.1016/j.cpc.2014.10.006},
pages = {20-28},
volume = {187}
}

@article{Voigtlander2024,
   author = {Anne Voigtländer and Morgane Houssais and Karol A. Bacik and Ian C. Bourg and  Justin C. Burton and Karen E. Daniels and Sujit S. Datta and Emanuela Del Gado and Nakul S. Deshpande and Olivier Devauchelle and Behrooz Ferdowsi and Rachel Glade and Lucas Goehring and Ian J. Hewitt and Douglas Jerolmack and Ruben Juanes and Arshad Kudrolli and Ching-Yao Lai and Wei Li and Claire Masteller and Kavinda Nissanka and Allan M. Rubin and Howard A. Stone and Jenny Suckale and  Nathalie M. Vriend and John S. Wettlaufer and Judy Q. Yang},
   title = {Soft matter physics of the ground beneath our feet},
 journal = {Soft Matter},
year = {2024},
doi = {https://doi.org/10.1039/D4SM00391H},
pages = {5859-5888)},
volume = {20}
}

@article{Guo2023,
   author = {Xingsen Guo and Thorsten Stoesser and Defeng Zheng and Qianyu Luo and Xiaolei Liu and Tingkai Nian},
   title = {A methodology to predict the run-out distance of  submarine landslides},
 journal = {Computers and Geotechnics},
year = {2023},
doi = {https://doi.org/10.1016/j.compgeo.2022.105073},
pages = {1-14},
volume = {153}
}

@article{Lee2019,
   author = {Cheng-Hsien Lee},
   title = {Multi-phase flow modeling of submarine landslides: Transformation from hyperconcentrated flows into turbidity currents},
 journal = {Advances in Water Resources},
year = {2019},
pages= {1-8},
volume = {131},
doi={10.1016/j.advwatres.2019.103383}
}

@article{Soulaine2024,
   author = {Cyprien Soulaine},
   title = {Micro‐Continuum Modeling: An Hybrid‐Scale Approach for Solving Coupled Processes in Porous Media},
 journal = {Water Resources Research},
year = {2024},
pages= {1-20},
volume = {60},
doi ={https://doi.org/10.1029/2023WR035908}
}

@article{Clarke1990,
   author = {John Hughes Clarke and Alexander Shor and David Piper and Larry Mayer },
   title = {Large‐scale current‐induced erosion and deposition in the path of the 1929 Grand Banks turbidity current},
 journal = {Sedimentology},
year = {1990},
pages= {613-629},
volume = {37},
doi ={https://doi.org/10.1111/j.1365-3091.1990.tb00625.x}
}

@article{Mewis2009,
   author = {Jan Mewis and Norman J. Wagner},
   title = {Thixotropy},
 journal = {Advances in colloid and interface science},
year = {2009},
pages= {214-227},
volume = {147-148},
doi ={https://doi.org/10.1016/j.cis.2008.09.005}
}

@article{Pudasaini2012,
   author = {Shiva P. Pudasaini},
   title = {A general two‐phase debris flow model},
 journal = {Journal of Geophysical Research: Earth Surface},
year = {2012},
pages= {1-28},
volume = {117},
doi ={https://doi.org/10.1029/2011JF002186}
}

@article{Paull2018,
   author = {Charles K. Paull and Peter J. Talling and Katherine L. Maier and Daniel Parsons and Jingping Xu and David W. Caress and Roberto Gwiazda and Eve M. Lundsten and Krystle Anderson and James P. Barry and Mark Chaffey and Tom O’Reilly and Kurt J. Rosenberger and Jenny A. Gales and Brian Kieft and Mary McGann and Steve M. Simmons and Mike McCann and Esther J. Sumner and Michael A. Clare and Matthieu J. Cartigny},
   title = {Powerful turbidity currents driven by dense basal layers},
 journal = {Nature communications},
year = {2018},
pages= {1-9},
volume = {9},
doi ={https://doi.org/10.1038/s41467-018-06254-6}
}

@article{Sherwood2022,
   author = {Christopher R. Sherwood and Ap van Dongeren and James Doyle and Christie A. Hegermiller and Tian-Jian Hsu and Tarandeep S. Kalra and Maitane Olabarrieta and Allison M. Penko and Yashar Rafati and Dano Roelvink and Marlies van der Lugt and Jay Veeramony and John C. Warner},
   title = {Modeling the morphodynamics of coastal responses to extreme events: What shape are we in?},
 journal = {Annual review of marine science},
year = {2022},
pages= {457-492},
volume = {14},
doi ={https://doi.org/10.1146/annurev-marine-032221-090215}
}

@article{Pignon1998,
   author = {Frédéric Pignon and Albert Magnin and Jean-Michel Piau},
   title = {Thixotropic behavior of clay dispersions: combinations of scattering and rheometric techniques},
 journal = {Journal of Rheology},
year = {1998},
pages= {1349-1373},
volume = {42},
doi ={https://doi.org/10.1122/1.550964}
}

@article{Berend1995,
   author = {Isabelle Bérend and 
Jean-Maurice Cases and Michèle François and Jean-Pierre Uriot and Laurent Michot and Armand Masion and Fabien Thomas},
   title = {Mechanism of adsorption and desorption of water vapor by homoionic montmorillonites: 2. The Li+, Na+, K+, Rb+ and Cs+-exchanged forms},
 journal = {Clays and Clay Minerals},
year = {1995},
pages= {324-336},
volume = {43},
doi ={https://doi.org/10.1346/CCMN.1995.0430307}
}

@article{Hage2019,
   author = {Sophie Hage and Matthieu J.B. Cartigny and Esther J. Sumner and Michael A. Clare and John E. Hughes Clarke and Peter J. Talling and D. Gwyn Lintern and Stephen M. Simmons and Ricardo Silva Jacinto and Age J. Vellinga and Joshua R. Allin and Maria Azpiroz‐Zabala and Jenny A. Gales and Jamie L. Hizzett and James E. Hunt and Alessandro Mozzato and Daniel R. Parsons and Ed L. Pope and Cooper D. Stacey and William O. Symons and Mark E. Vardy and Camilla Watts},
   title = {Direct monitoring reveals initiation of turbidity currents from extremely dilute river plumes},
 journal = {Geophysical Research Letters},
year = {2019},
pages= {11310-11320},
volume = {46},
doi ={https://doi.org/10.1029/2019GL084526}
}

@article{Hampton1972,
   author = {Monty A. Hampton},
   title = {The role of subaqueous debris flow in generating turbidity currents},
 journal = {Journal of Sedimentary Research},
year = {1972},
pages= {775-793},
volume = {42}
}

@article{Vowinckel2023,
   author = {Bernhard Vowinckel and Kunpeng Zhao and Rui Zhu and Eckart Meiburg},
   title = {Investigating cohesive sediment dynamics in open waters via grain-resolved simulations},
 journal = {Flow},
year = {2023},
pages= {1-41},
volume = {3},
doi ={https://doi.org/10.1017/flo.2023.20}
}

%
%
%
%
%

\end{document}